\documentclass[12pt,preprint2]{aastex61}

\usepackage{natbib}
\usepackage{graphicx}

\usepackage{bmpsize}
\usepackage{floatrow}
\usepackage{changepage}
\usepackage{mathtools}
\usepackage{amsmath}
\usepackage{appendix}
\usepackage{url}
\DeclareUnicodeCharacter{00B4}{}

\shorttitle{Estimating Host Prop.}
\shortauthors{Coleman et al.}

\begin{document}

\title{Accretion History of AGN: Estimating the Host Galaxy Properties in X-ray Luminous AGN from z=0-3}

\author{Brandon Coleman}
\affiliation{Department of Physics \& Astronomy, University of Kansas,
Lawrence, KS 66045, USA}

\author{Allison Kirkpatrick}
\affiliation{Department of Physics \& Astronomy, University of Kansas,
Lawrence, KS 66045, USA}

\author{Kevin C. Cooke}
\affiliation{Department of Physics \& Astronomy, University of Kansas,
Lawrence, KS 66045, USA}

\author{Eilat Glikman}
\affiliation{Department of Physics, Middlebury College, Middlebury,
VT 05753, USA}

\author{Stephanie La Massa}
\affiliation{Space Telescope Science Institute, 3700 San Martin Dr, Baltimore, MD 21218, USA}

\author{Stefano Marchesi}
\affiliation{INAF - Osservatorio di Astrofisica e Scienza dello Spazio di Bologna, Via Piero Gobetti, 93/3, 40129, Bologna, Italy}

\author{Alessandro Peca}
\affiliation{Physics Department, University of Miami, Coral Gables, FL 33155, USA}

\author{Ezequiel Treister}
\affiliation{Instituto de Astrof\'{i}sica, Facultad de F\'{i}sica, Pontificia Universidad Cat´olica de Chile, Casilla 306, Santiago 22, Chile}

\author{Connor Auge}
\affiliation{Institute for Astronomy, University of Hawaii, 2680 Woodlawn Drive, Honolulu, HI 96822, USA}

\author{C. Megan Urry}
\affiliation{Yale Center for Astronomy \& Astrophysics, New Haven, CT 06520, USA}
\affiliation{Department of Physics, Yale University, PO BOX 201820, New Haven, CT 06520, USA}

\author{Dave Sanders}
\affiliation{Institute for Astronomy, University of Hawaii, 2680 Woodlawn Drive, Honolulu, HI 96822, USA}

\author{Tracey Jane Turner}
\affiliation{Eureka Scientific, Inc., 2452 Delmer Street, Suite 100, Oakland, CA 94602-3017, USA}

\author{Tonima Tasnim Ananna}
\affiliation{Yale Center for Astronomy \& Astrophysics, New Haven, CT 06520, USA}
\affiliation{Department of Physics, Yale University, PO BOX 201820, New Haven, CT 06520, USA}

\begin{abstract}
We aim to determine the intrinsic far-Infrared (far-IR) emission of X-ray-luminous quasars over cosmic time. Using a 16 deg$^2$ region of the Stripe 82 field surveyed by \textit{XMM-Newton} and \textit{Herschel Space Observatory}, we identify 2905 X-ray luminous ($L_X > 10^{42}$ erg/s) Active Galactic Nuclei (AGN) in the range $z \approx 0-3$.
The IR is necessary to constrain host galaxy properties such as star formation rate (SFR) and gas mass. However, only 10\% of our AGN are detected both in the X-ray and IR. Because 90\% of the sample is undetected in the far-IR by \textit{Herschel}, we explore the mean IR emission of these undetected sources by stacking their \textit{Herschel}/SPIRE images in bins of X-ray luminosity and redshift. We create stacked spectral energy distributions from the optical to the far-IR, and estimate the median star formation rate, dust mass, stellar mass, and infrared luminosity using a fitting routine. We find that the stacked sources on average have similar SFR/L$_{bol}$ ratios as IR detected sources. The majority of our sources fall on or above the main sequence line suggesting that X-ray selection alone does not predict the location of a galaxy on the main sequence. We also find that the gas depletion timescales of our AGN are similar to those of dusty star forming galaxies. This suggests that X-ray selected AGN host high star formation and that there are no signs of declining star formation. 
\end{abstract}

\keywords{Galaxies:active, Galaxies:evolution, Galaxies:nuclei Galaxies:quasars:supermassive black holes, Galaxies:Seyfert, galaxies:star formation}

\section{Introduction}
Quasars, which can be the most luminous non-transient objects in the known 
Universe, can be created through major mergers as one possible scenario. When two gas-rich disk galaxies collide and coalesce, the merger fuels a rapid burst of star formation and triggers a 
period of rapid growth of a supermassive black hole (SMBH) \citep{1988ApJ...328L..35S,2006ApJ...652..864H}. 
In this scenario, the quasar experiences an obscured phase, when the optically bright accretion disk fueling the SMBH is enshrouded by a thick layer of gas and dust, also known as the torus, while its host galaxy experiences higher star formation, also triggered by the merger \citep{2017MNRAS.468.1273R,2010Sci...328..600T,perna2018}. The obscured phase is followed by the blowout phase when the quasar launches winds which expel the obscuring dust. The blowout makes the accretion disk visible in the optical again, and the host galaxy begins to quench its star formation \citep{2004ApJ...607...60G,2012ApJ...757...51G}. In the major merger scenario, the heavily obscured phases are likely short lived \citep{2012ApJ...757...51G}. Through clustering measurements in \citet{2013ApJ...762...70C} and \citet{2016ApJ...828...90L}, the lifetime of quasars is weakly constrained to be between  $10^6 - 10^9$ yr.

The major merger scenario suggests a connection between black hole growth and star formation, mainly through negative feedback quenching star formation. However, this link is observationally difficult to identify.

For example, \citet{2017MNRAS.472.2221S} show no decrease in star formation rate (SFR) with AGN bolometric luminosity for a given redshift using optically-selected AGN from the Sloan Digital Sky Survey (SDSS). \citet{2019MNRAS.488.1180S} found that a population of 20 unobscured quasars at $z = 2$ have SFRs that are similar to those of star-forming galaxies (SFGs) based on ALMA continuum measurements. \citet{2015MNRAS.453L..83M} claims the existence of a significant AGN population below the main-sequence of star-forming galaxies. On the other hand, \citet{2021ApJ...910..124X} find quasars with elevated SFRs that lie above the main sequence.

\par
Quasars are typically separated into ``red" (obscured) and ``blue" (unobscured) classifications, based on their optical/near-IR colors.
\citet{kirkpatrick2020} identified a unique population of quasars, dubbed ``cold quasars''. These objects do not fit neatly into the red or blue quasar populations. Cold quasars are unobscured, quantified through optical colors and the detection of broad lines in their spectra, that have a large amount of cold dust present quantified through far-IR emission. This cold dust allows these quasars to be detected in all three \textit{Herschel}/SPIRE bands (250, 350, and 500 $\mu$m). The cold quasar sample exhibits intense SFRs in comparison to normal star-forming galaxies, again calling into question whether quasar feedback linked with declining star formation. \par

\par
Studying the properties of the quasar host galaxies, particularly their SFRs, is difficult. Selection at different wavelengths can produce biased samples, especially with obscured quasars \citep{2009ApJ...696..891H,2017ApJ...835...27A}. Unobscured quasars are luminous and outshine their hosts both in the optical and mid-IR making characterization of the hosts difficult \citep{2011ApJ...733...31H,2012ApJ...753...30S}. The optical $H\alpha$ emission line  frequently used to measure star formation is contaminated by the quasar emission \citep{1981PASP...93....5B,2012MNRAS.420.1061T}. The submillimeter emission lines that are used to trace the interstellar medium in star forming regions can also be contaminated by the quasar \citep{kirkpatrick2019,2016ApJ...822L..10I}. Therefore, the most robust tracer of the host galaxy is the far-IR, since quasars are expected not to substantially heat dust beyond a wavelength of greater than 100 $\mu$m, or roughly 30 K \citep{2011MNRAS.414.1082M}. The far-IR regime allows us to determine parameters overcoming the aforementioned biases such as SFR, IR luminosity, and dust mass that describe the host galaxy. However, the limited lifespan, resolution, and sensitivity of far-IR observatories create difficulties in acquiring the necessary observations of quasars. For example, ALMA can only trace the Rayleigh Jeans tail of the IR emission. Therefore, it makes it difficult for observers to determine the AGN's contribution to the far-IR emission. Until the launch of a new far-IR telescope, we must rely on IR stacking to determine the average properties of galaxies too faint to be detected individually.
\par

In this paper, we look at parameters that describe the far-IR properties of X-ray bright, but IR-undetected active galactic nuclei (AGN). We distinguish between AGN ($L_X<10^{44}$\,erg/s) and quasars ($L_X\geq 10^{44}$\,erg/s) for clarity when comparing to other populations. We note that IR-undetected AGN in this paper will be the AGN that are not detected by the
\textit{Herschel Space Telescope}, although they do have
WISE detections, indicating a significant amount of hot
dust. We seek to compare the host galaxy properties of IR-undetected AGN with that of different samples and types of AGN to obtain a better picture of the evolutionary timeline of such objects and to constrain the average host galaxy properties of unobscured quasars. We use the cold quasar sample in this paper to make comparisons to confirm their intense host galaxy properties. In Section \ref{sec:Data}, we discuss the
Accretion History of AGN survey (AHA; PI M. Urry) from which we select our sample. We also discuss our stacking method for Herschel undetected AGN, and the SED fitting procedure, {\sc Sed3Fit}, used for our sample. We discuss in Section \ref{sec:Results} how SFR, dust mass, stellar mass, and IR luminosity of our sample compare with that of other obscured and unobscured quasars. In Section \ref{sec:Discussion}, we discuss possible evolutionary scenarios of our unobscured quasars, and in Section \ref{sec:conclusions} we present our conclusions. Throughout the paper, we assume a standard cosmology with $H_0 = 70 \;km \;s^{-1}\; Mpc^{-1}, \Omega_M = 0.3$ and $\Omega_{\Lambda} = 0.7$.

\section{Data and Stacking}\label{sec:Data}
To uncover a significant sample of high-luminosity AGN, large volumes of the Universe must be surveyed. We use data from the Stripe~82X survey \citep{2013MNRAS.432.1351L,2013MNRAS.436.3581L,2015ApJ...800..144L}, part of the multi-wavelength survey Accretion History of AGN (AHA, PI M. Urry). Due to the large volume surveyed, the Stripe~82X survey contains the most X-ray luminous sources compared to other X-ray surveys like COSMOS and GOODS-South.
\par
The AHA survey (S82X) covers a total of 31.3 deg$^2$ of the sky in the Stripe 82 region, consisting of 4.6 deg$^2$ and 15.6 deg$^2$ from targeted observations with \textit{XMM-Newton} in AO10 and AO13 \citep{2013MNRAS.436.3581L,2016ApJ...817..172L} combined with 5.6 deg$^2$ of \textit{XMM-Newton} archival pointings and 6.0 deg$^2$ of \textit{Chandra X-ray Observatory} archival pointings. In this paper, we focus on the 15.6 deg$^2$ \textit{XMM-Newton} observations that are fully overlapped by infrared observations from the SPIRE instrument on the \textit{Herschel Space Observatory} as part of the Herschel Stripe 82 Survey (HerS; \citet{2014ApJS..210...22V}). The HerS survey has a $3\sigma$ detection limit of $S_{250 \mu m} = 30\; mJy$ \citep{2014ApJS..210...22V}. 

Our parent sample contains 3200 X-ray sources. Out of this sample, about 70\% have spectroscopic redshifts. For the other 30\%, the photometric redshifts were calculated use the \texttt{Le PHARE} (Photometric Analysis for Redshift Estimation) code \citep[see][for details of the photo-$z$ calculation]{2017ApJ...850...66A}. For the sources with both photo-$z$s and spec-$z$s, approximately 14\% have $\Delta z /(1+z_{\rm spec}>0.15$. 
We discuss the potential impact of these overestimations in Section \ref{sec:stack}.  Figure 1 in \citet{2017ApJ...850...66A} shows the layout of Stripe 82 covered by \textit{XMM-Newton}.

\par
Multiwavelength counterpart matching was done in \citet{2016ApJ...817..172L} and \citet{2017ApJ...850...66A}. The authors match the X-ray positions to counterparts spanning the UV to the mid-IR. Both papers used ancillary data from the co-added Sloan Digital Sky Survey (SDSS) catalogs \citep{2014ApJS..213...12J,2016MNRAS.456.1359F}, which reach 2.5 mag deeper than single epoch data. This increased the likelihood of each X-ray source having an optical counterpart.  The AO13 region of the survey is also covered by \textit{Spitzer Space Telescope} (in 3.6 \& 4.5 $\mu$m) through the \textit{Spitzer}/HETDEX Exploratory Large Area (SHELA) survey \citep{2016ApJS..224...28P} and the \textit{Spitzer} IRAC Equatorial Survey (SpIES; \citet{2016ApJS..225....1T}). The entire stripe is also covered by \textit{WISE} in the mid-IR and the Vista Hemisphere Survey in the near-IR \citep{2013Msngr.154...35M}. We would like to note that multiwavelength matching is outside of the bounds of this paper. For more information on the process of multiwavelength counterpart matching, please see \citet{2017ApJ...850...66A}.

\subsection{The Herschel Subsample \& Cold Quasars}
One of the goals of AHA is to measure  properties of the host galaxy such as SFR and IR luminosity in the most luminous X-ray sources to investigate the interplay between star formation and black hole growth. The far-IR is critical because in unobscured luminous AGN, the host galaxy can be outshone at every other wavelength \citep{2006ApJS..166..470R}. However, only 120 of our X-ray sources are detected at 250 $\mu$m. We refer to these sources as the Herschel subsample.  Thirty of these sources meet our criteria for being unobscured quasars with $L_X > 10^{44}$ erg/s, $M_B < -23$, and broad emission lines in their optical spectra \citep{kirkpatrick2020}. Note that the X-ray luminosities are observed values, calculated directly using observed fluxes. Due to the extreme infrared luminosities ($L_{\rm IR} > 10^{11} L_\odot$), we refer to these sources as cold quasars because they still contain a large reservoir of cold dust leading to high SFRs compared to star forming galaxies \citep{kirkpatrick2020}. The X-ray luminosities have not been corrected for obscuration fraction. However, Figure \ref{obscure} shows that the X-ray emission in our sample on average is unobscured, which makes the obscuration fraction negligible. For the remaining sources with no clear IR detection, it is more difficult to characterize the host galaxy. 

\subsection{Herschel-Undetected AGN}\label{sec:stack}
The AO13 region of Stripe 82X that overlaps with the HerS survey area contains approximately 3195 X-ray sources \citep{2016ApJ...817..172L}. We remove 388 sources that do not have any robust multi-wavelength counterparts, as indicated by a counterpart quality of 3 or 4 in the \citet{2017ApJ...850...66A} photometric catalog. This means that the source has different counterparts in multiple bands with comparable likelihood ratios, but have different reliability classes. This does not qualify as robust. Also, a quality flag of 4 means there is a counterpart in only one band, which does not qualify as multi-wavelength counterparts. We further limit the sample to galaxies with a detection in the WISE W1 band. These are the galaxies most likely to contain significant dust emission, and the mid-IR coverage of WISE is necessary to accurately constrain the AGN emission. We remove 965 sources which have no WISE W1 detection as they will not have a significant dust emission present and skew the SED fitting process. This leaves us with 1842 sources, 120 of which form the Herschel sample and are not included in the infrared stacking. All removed sources are evenly distributed over $z \approx 0-3$ and $L_X=10^{42}-10^{46}$erg/s, so their removal does not present a systematic bias in either of these parameters.

\begin{figure}
	\includegraphics[width=3.4in]{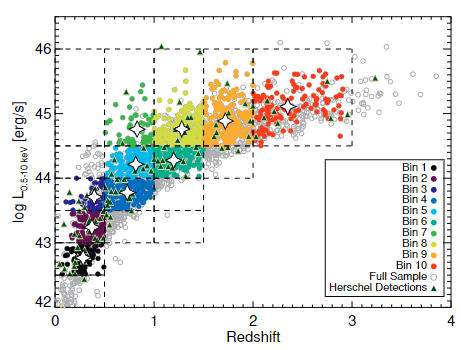}
    \caption{The X-ray Luminosity and redshift distribution of our sample and parent population. The grey dots are every galaxy in AHA \citep{2017ApJ...850...66A} that have a multiwavelength counterpart and lie in the region of Stripe82 covered by HerS \citep{2014ApJS..210...22V}. We subdivided sources by $L_X$ and $z$ (dashed lines) and then stacked Herschel/SPIRE images in bins with more than 50 sources. The colored points show which sources we stacked--these sources had a WISE W1 detected but lacked a SPIRE 250\,$\mu$m detection. We also overplot the 120 sources (green triangles) that are detected at 250\,$\mu$m.}
    \label{lxvz}
\end{figure}

\begin{figure}
	\includegraphics[width=3.4in]{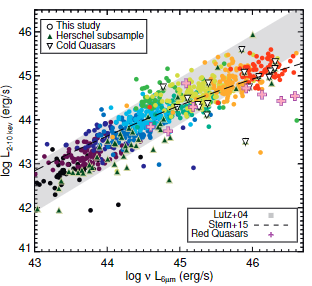}
    \caption{Hard X-ray luminosity as a function of 6 $\mu$m luminosity. We plot the relationship derived from local galaxies (gray shaded region)\citep{2004A&A...418..465L}, the relationship derived from higher luminosity, higher redshift sources (dashed line)\citep{2015ApJ...807..129S}, red quasars, the cold quasars, and our stacked sample. The colors of the stacked sources match those of Figure \ref{lxvz}. This plot tells us that the X-ray emission we observe in our sources are similar to that of local galaxies resulting in unobscured X-ray emission.}
    \label{obscure}
\end{figure}

\begin{figure*}
\centering
\includegraphics[width=2in]{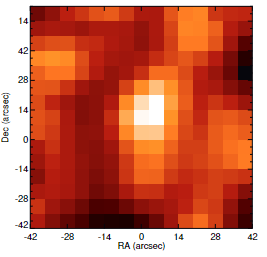}
\includegraphics[width=2in]{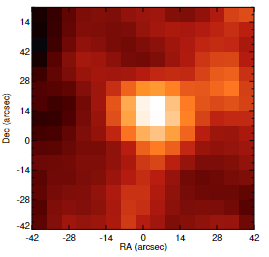}
\includegraphics[width=2in]{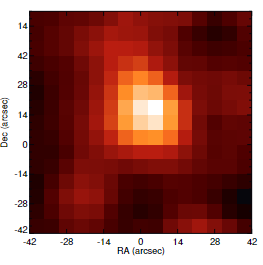}
\includegraphics[width=2in]{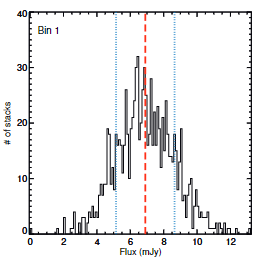}
\includegraphics[width=2in]{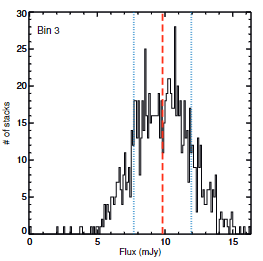}
\includegraphics[width=2in]{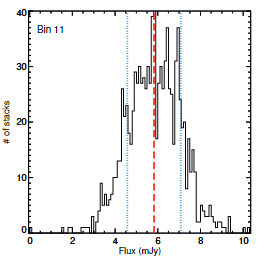}
\caption{We show the results of our stacking method for three representative bins (Bin 1, {\it left}; Bin 3, {\it middle}; Bin 10{\it right}). The top panels show the 250\,$\mu$m image resulting from 1000 bootstrapped stacks. The bottom panels show the resulting histograms, with the mean (red solid line) and standard deviations (blue dashed lines) marked.}
\label{fig:stack}
\end{figure*}

The far-IR regime is observable using \textit{Herschel}/SPIRE bandpasses at 
250, 350, and 500 $\mu$m. Although there is no detection for 1722 sources in the \textit{Herschel}/SPIRE bands, we use the SPIRE maps to obtain a mean flux for the Herschel non-detected sample. We separated our sources into bins first by the 0.5-10 keV X-ray luminosity.  We split sources into luminosity bin sizes 
of 0.5\,dex, from $\log L_X=42-46$\,[erg/s]. We further subdivide by redshift in bins of $\Delta z =0.5$ for $z=0-2$ and then one high redshift bin spanning $z=2-3$. We only perform stacking in bins with more than 50 sources, as below this threshold, we are unable to obtain a 3$\sigma$ stacking detection. Even so, in the bin spanning $z=0.5-1, \log L_X=43.0-43.5\,$[erg/s], we were unable to obtain a detection at a SPIRE wavelength, so we omit this bin from further analysis. These selection criteria result in 10 bins, which are illustrated in Figure \ref{lxvz}. We show examples of the 250\,$\mu$m stacked images in Figure \ref{fig:stack}.

To stack, we follow the methodology outlined in \citet{2017MNRAS.472.2221S}, so that we can compare directly to that sample. We use the positional coordinates of the optical counterpart for each source, listed in the \citet{2017ApJ...850...66A} X-ray catalog. 
Because SPIRE images have units of Jansky/beam, it is straightforward to stack the central pixels at the location of each source. We do this for all sources in a given bin and take the mean. When we stack our sources, we must take into account the photometric redshifts and their impact. 
To determine the significance of overestimated redshifts, we estimate the uncertainties through bootstrapping each bin 1000 times. The results show little to no change in the overall stack and SED. The final reported flux and uncertainty is the mean and standard deviation of the bootstrap trials. We list these fluxes in Table \ref{fluxes}.

\begin{deluxetable*}{rrcccrcrcr}
\tabletypesize{\footnotesize}
\tablecolumns{10}
\tablewidth{0pt}

\tablecaption{Herschel Stacked Photometry}
\tablehead{
\colhead{Bin} & \colhead{$z$ range} & \colhead{$L_{0.5-10\,keV}$ range} & \colhead{No. of Sources}\tablenotemark{a} & $S_{250}$\tablenotemark{a} & $3\sigma_{250}$\tablenotemark{b} & $S_{350}$ & $3\sigma_{350}$\tablenotemark{b} & $S_{500}$ & $3\sigma_{500}$\tablenotemark{b} \\
 &  & \colhead{($\log$ erg/s)} &  &  (mJy) & (mJy) & (mJy) & (mJy) & (mJy) & (mJy)}
\startdata
1   & 0.0--0.5  & 42.5--43.0    & 83  [114]  & 6.77$\pm$1.72 [6.64] & 5.08  & \nodata  & 4.84   & \nodata & 5.44 \\
2   & 0.0--0.5  & 43.0--43.5    & 144 [182]  & 6.22$\pm$1.78 [6.03] & 3.82  & 4.44$\pm$1.35 & 3.74  & \nodata & 4.16\\
3   & 0.0--0.5  & 43.5--44.0    & 66  [81]  & 9.93$\pm$2.05 [9.52] & 5.61  & \nodata   & 5.46  & \nodata & 5.99 \\
4   & 0.5--1.0  & 43.5--44.0    & 214 [319]  & 3.81$\pm$0.97 [4.07] & 3.13  & \nodata   & 3.06  & \nodata & 3.41 \\
5   & 0.5--1.0  & 44.0--44.5    & 176 [231]  & 5.22$\pm$1.47 [5.91] & 3.50  & \nodata   & 3.45 & \nodata & 3.81  \\
6   & 1.0--1.5  & 44.0--44.5    & 169 [277]  & 6.88$\pm$1.29 [5.50] & 3.60  & 4.68$\pm$1.09 & 3.50 & \nodata & 3.91  \\
7   & 0.5--1.0  & 44.5--46.0    & 58  [64]  & 6.78$\pm$1.85 [9.57] & 5.96  & \nodata   & 5.84  & \nodata & 6.48   \\
8   & 1.0--1.5  & 44.5--46.0    & 191 [256]  & 6.77$\pm$1.16 [6.72] & 3.33  & 5.63$\pm$1.01 & 3.23  & \nodata & 3.63   \\
9  & 1.5--2.0  & 44.5--46.0    & 223 [348]  & 6.00$\pm$1.08 [4.33] & 3.09  & 6.29$\pm$0.91 & 3.04  & 4.99$\pm$1.02 & 3.36 \\
10  & 2.0--3.0  & 44.5--46.0    & 135 [311]  & 5.81$\pm$1.20 [4.81] & 3.95  & 4.48$\pm$1.23 & 3.83  & \nodata & 4.29 \\
\enddata
\tablenotetext{a}{Listed in brackets are the number of sources the stacked $S_{250}$ flux if the galaxies without WISE detections are included.}
\tablenotetext{b}{3$\sigma$ detection limits were determined through stacking at random positions within the SPIRE maps 10,000 times and calculating the standard deviation of the results. If a bin is not detected, it's stacked flux falls below this limit.}
\label{fluxes}
\end{deluxetable*}

We determine which fluxes are significant detections by stacking at random positions within the SPIRE maps. Following \citet{2017MNRAS.472.2221S}, we do this separately for each bin, using the number of sources within each bin as the number of random positions to select in the SPIRE image. We stack 10,000 times for each bin and calculate the mean and standard deviation. The mean represents the bias due to the confusion limit, which can be significant at far-IR wavelengths. However, we find that in general, $\mu_{250}=0.16$\,mJy, $\mu_{350}=0.13$\,mJy, and $\mu_{500}=0.35$\,mJy, which is not significantly elevating our stacked fluxes. We consider three times the standard deviation to be the detection limit for each bin, and we list these 3$\sigma$ values in Table \ref{fluxes}. If a bin does not reach the 3$\sigma$ level, then we do not report the flux and instead use a 3$\sigma$ upper limit in our analysis. At 500\,$\mu$m, all bins except one are undetected. It is possible there is a bias on our fluxes due to clustering, although we expect any contamination to be at the 10\% level \citep{wang2015}.
\par

Finally, we test what effect removing the non-WISE detected sources from our sample has on the stacked fluxes by restacking the 250\,$\mu$m images with the WISE and non-WISE sources. We list the results in brackets in Table \ref{fluxes}. In most cases, including the non-WISE sources has no significant effect. In the three most luminous bins, the non-WISE sources significantly lower the stacked flux. Bin 7 has a significant increase in flux, likely due to one or two anomalous sources.

We created mean multiwavelength spectral energy distributions (SEDs) for each bin. We calculated the mean flux in each filter from the UV to the mid-IR. In some cases, galaxies would be missing photometry in a given band. So as not to bias our data, we created mock photometry when a value was missing. \citet{2017ApJ...850...66A} determined photometric redshifts for all galaxies in AHA by fitting a suite of AGN and quasar templates. We use the best-fit template, listed in the \citet{2017ApJ...850...66A} catalog, scaled to the available photometry. We then convolved with the appropriate transmission curve to estimate the flux. We included these fluxes along with the detected fluxes. We again followed a bootstrap technique and calculated the mean flux in each bandpass 1000 times, resampling with replacement each time. The final reported flux is the mean of the bootstrap trials, and the uncertainty is the standard deviation. We list the photometry for each bin in Table \ref{photometry} in Appendix \ref{app:A}.
\par
The use of mock photometry when it is missing in a particular band for a particular source may introduce additional uncertainties. To determine the significance of using mock photometry, we execute the stacking process as before. However, for this scenario we stack the sources in each bin for which there are measurements in each specific band. In Table \ref{nomock} in Appendix \ref{app:A} we state the number of sources in each bin that have a detection in each band. After grouping the new stacks into bins, we determine the average flux of the stack and compare it to the photometry obtained from mock photometry. Comparing the errors in Table \ref{photometry} to the flux difference values in Table \ref{nomock}, we see that when we exclude the mock photometry values, we observe fluxes that are on average within the errors of the fluxes obtained using mock photometry. We note that in Table \ref{nomock}, all sources in a particular bin were detected in some of the filters. In other words, no mock photometry was used for these filter-bin combinations and thus the change in flux is zero. This, in addition to the results of bootstrapping shows that using mock photometry has no significant effect to the produced photometry.
\par
We stacked all of the photometry in the observed frame of the sources, because the SPIRE stacking is by necessity done in the observed frame. We then converted to the rest frame using the mean redshift of the sources in the bin.  
The redshifts are listed in Table \ref{props}.

\begin{deluxetable*}{lccccccccc}
\tabletypesize{\footnotesize}
\tablecolumns{12}
\tablewidth{0pt}
\tablecaption{Stacked SED Properties}
\tablehead{
\colhead{Bin} & \colhead{mean $L_{X}$ ($\sigma$)} & \colhead{mean $z$ ($\sigma$)} & \colhead{$L_{bol}$} & \colhead{ $L_{IR}$} & \colhead{$M_\ast$} & \colhead{$M_{dust}$} & \colhead{SFR} & \colhead{AGN ctr.} & \colhead{Reduced $\chi^2$}\\
  & \colhead{($\log$ erg/s)} & & \colhead{($\log$ erg/s)} & \colhead{($\log L_{\odot}$)}& \colhead{($\log M_\odot$)} & \colhead{($\log M_\odot$)}
 & \colhead{($M_\odot /yr$)} &\colhead{to L$_{70-100 \mu m}$}& }
\startdata
Bin 1 & 42.77 (0.13) & 0.28 (0.09) & 44.26 & 11.20 $\pm$ 0.25 & 11.39 $\pm$ 0.38 & 7.12 $\pm$ 0.17 & 23.16 $\pm$ 10.21 & 0.041 & 0.81 \\ 
Bin 2 & 43.24 (0.14) & 0.37 (0.08) & 44.79 & 11.30 $\pm$ 0.09 & 11.68 $\pm$ 0.90 & 7.57 $\pm$ 0.28 & 8.51 $\pm$ 3.66 & 0.374 & 0.28 \\
Bin 3 & 43.71 (0.15) & 0.40 (0.09) & 45.32 & 11.17 $\pm$ 0.10 & 11.47 $\pm$ 0.14 & 7.88 $\pm$ 0.28 & 20.54 $\pm$ 9.4 & 0.01 & 0.57 \\
Bin 4 & 43.78 (0.14) & 0.73 (0.13) & 45.42 & 11.51 $\pm$ 0.02 & 11.18 $\pm$ 0.18 & 8.49 $\pm$ 0.18 & 8.79 $\pm$ 0.42 & 0.005 & 3.51 \\
Bin 5 & 44.22 (0.13) & 0.81 (0.12) & 45.91 & 11.75 $\pm$ 0.04 & 10.88 $\pm$ 0.21 & 7.86 $\pm$ 0.18 & 64.84 $\pm$ 9.01 & 0.011 & 2.97 \\
Bin 6 & 44.28 (0.14) & 1.20 (0.14) & 45.98 & 12.05 $\pm$ 0.05 & 10.05 $\pm$ 0.04 & 8.49 $\pm$ 0.29 & 125.2 $\pm$ 10.5 & 0.003 & 1.95 \\
Bin 7 & 44.76 (0.23) & 0.83 (0.13) & 46.47 & 12.15 $\pm$ 0.05 & 12.06 $\pm$ 0.80 & 8.06 $\pm$ 0.35 & 340.5 $\pm$ 41.4 & 0.003 & 0.27 \\
Bin 8 & 44.76 (0.21) & 1.28 (0.14) & 46.50 & 12.25 $\pm$ 0.08 & 10.09 $\pm$ 0.42 & 8.71 $\pm$ 0.18 & 21.82 $\pm$ 11.46 & 0.001 & 1.08 \\
Bin 9 & 44.88 (0.27) & 1.72 (0.14) & 46.63 & 12.19 $\pm$ 0.06 & 11.22 $\pm$ 0.34 & 8.66 $\pm$ 0.17 & 85.05 $\pm$ 25.05 & 0.004 & 0.42 \\
Bin 10 & 45.11 (0.27) & 2.35 (0.27) & 46.91 & 12.58 $\pm$ 0.10 & 11.12 $\pm$ 0.14 & 8.65 $\pm$ 0.35 & 753.6 $\pm$ 142.9 & 0.004 & 0.52 \\
\enddata
\tablecomments{The uncertainties in X-ray luminosity and redshift are listed in parentheses and are derived from the distribution of sources in each bin.}
\label{props}
\end{deluxetable*}

\subsection{Fitting SEDs with {\sc Sed3Fit}}
We perform SED decomposition on our photometry with the SED-fitting code {\sc Sed3Fit} \citep{2013A&A...551A.100B}  
to estimate parameters such as dust mass, stellar mass, IR luminosity, IR AGN contribution, and star formation rate.
\par
{\sc Sed3Fit} performs SED fitting with a combination of three different components: stellar emission, dust emission from star formation, and a possible AGN accretion disk and torus. It does not include nebular emission lines. {\sc Sed3Fit} is a modified version of the fitting routine {\sc MAGPHYS} \citep{2008MNRAS.388.1595D} built to model AGN contribution, which {\sc MAGPHYS} does not natively include. The model libraries for the three components used in {\sc Sed3Fit} begin with stellar population emission spectra computed using the \citet{bruzual2003} or \citet{2007ASPC..374..303B} models including the effects of dust attenuation as prescribed by \citet{2000ApJ...539..718C}. The second component uses the emission of dust, computed by \citet{2008MNRAS.388.1595D}. Lastly, using the \citet{fritz2006} models, the AGN SEDs consists of isotropic emission of the central source. This emission consists of power laws in wavelengths of 0.001 - 20 $\mu$m. The torus models take into account a range of dust geometries and radii. Some of the emission of the AGN is either absorbed by the toroidal obscurer and re-emitted at IR wavelengths (1-1000 $\mu$m) or scattered by the same medium. The default set of models used in {\sc Sed3Fit} includes 10 torus models and randomly samples the optical/IR library 1000 times each. {\sc Sed3Fit} also uses a \citet{chabrier2003} initial mass function (IMF). 
\par
\begin{figure*}[ht]
    \centering
    \includegraphics[width=0.33\textwidth]{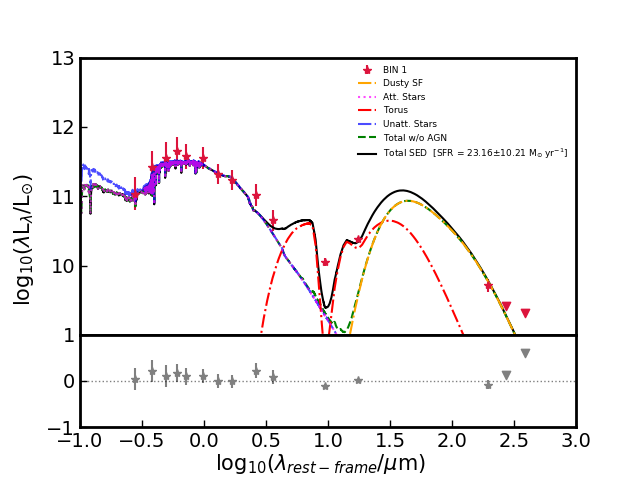}
    \includegraphics[width=0.33\textwidth]{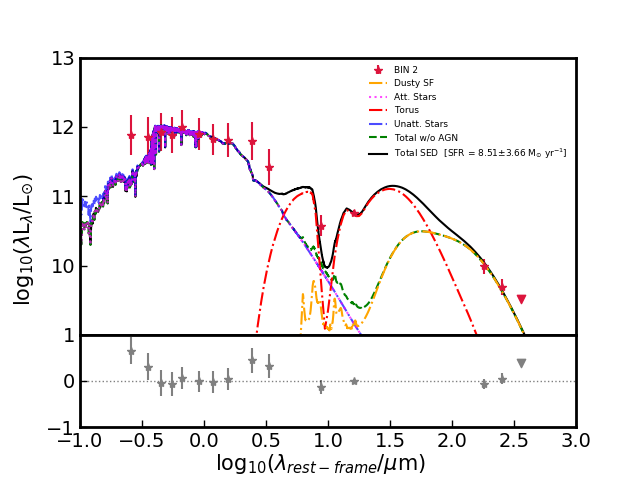}
    \includegraphics[width=0.33\textwidth]{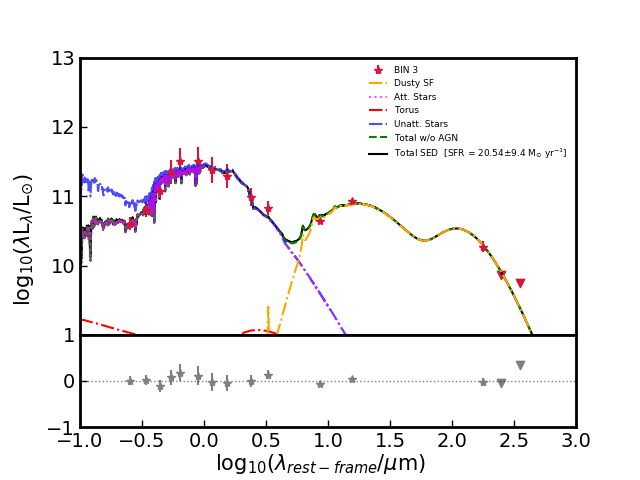}
    \includegraphics[width=0.33\textwidth]{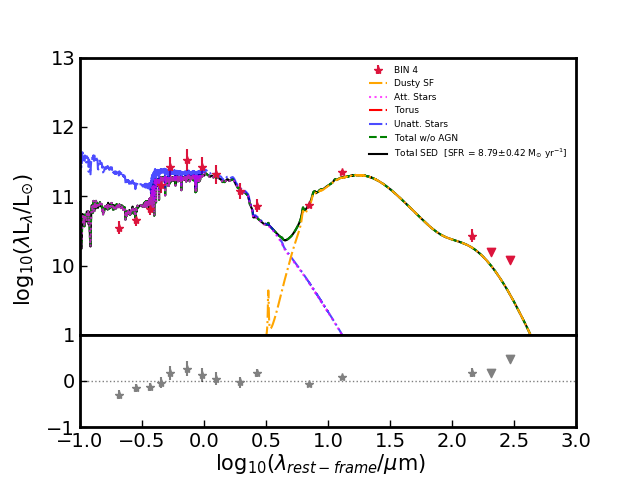}
    \includegraphics[width=0.33\textwidth]{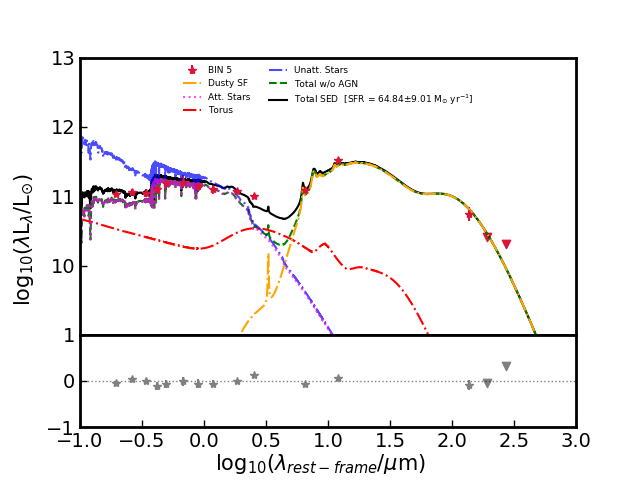}
    \includegraphics[width=0.33\textwidth]{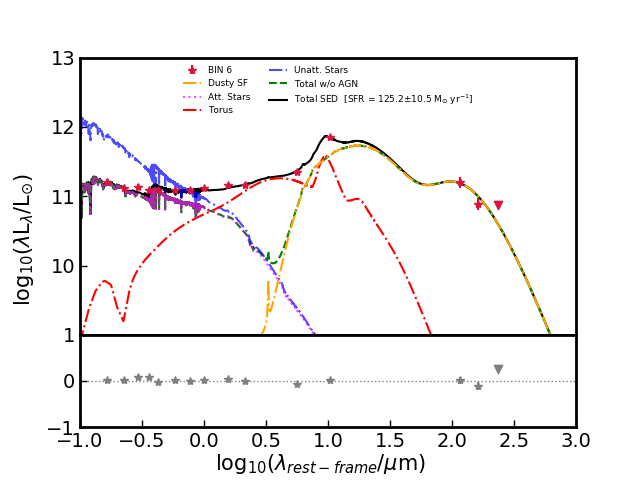}
    \includegraphics[width=0.33\textwidth]{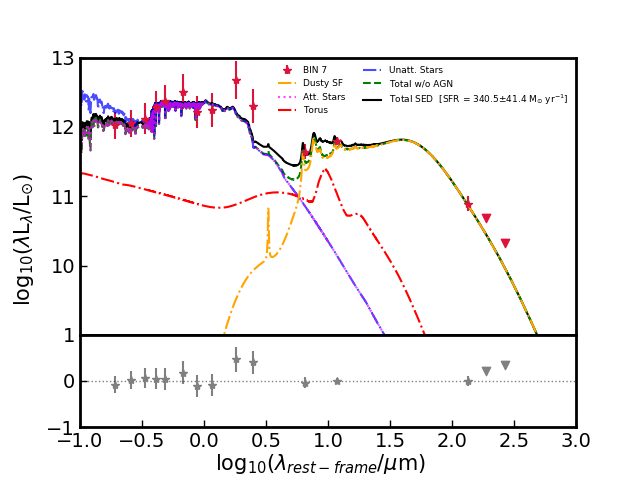}
    \includegraphics[width=0.33\textwidth]{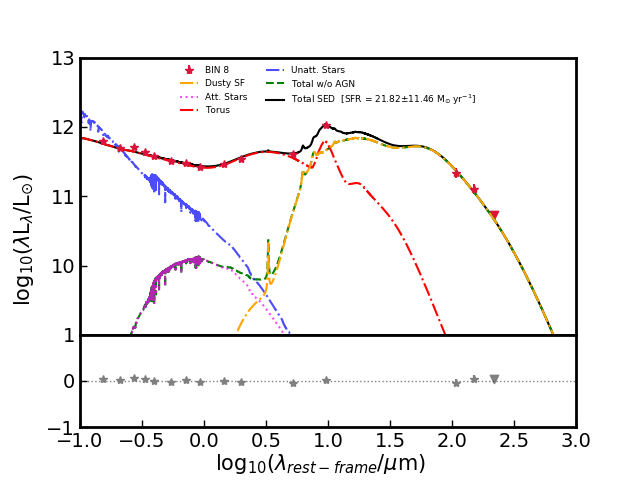}
    \includegraphics[width=0.33\textwidth]{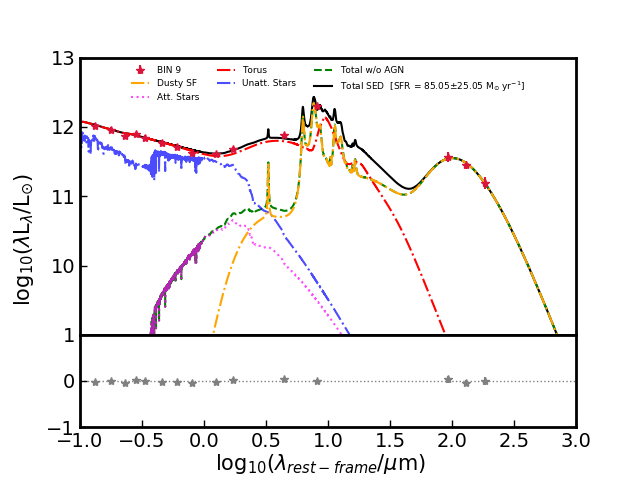}
    \includegraphics[width=0.33\textwidth]{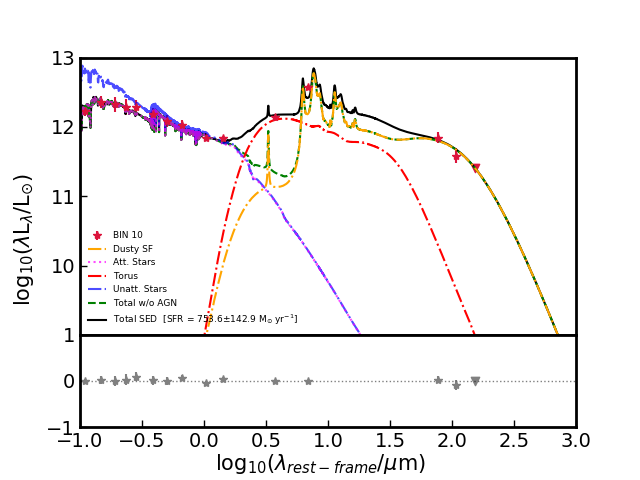}
    \caption{The SEDs created from using {\sc Sed3Fit}. Stars are the mean photometry and triangles indicate upper limits. Different components are represented by different colors. The SFR of the source is specified in the legend. We also plot the residuals of the observations with the best-fit SED model. \label{SEDfig}}
\end{figure*}

We performed fits for our 10 stacked SEDs, which are shown in Figure \ref{SEDfig}. 
We list the modeled properties $L_{IR,8-1000 \mu m}$, stellar mass (M$_\ast$), dust mass (M$_{dust}$), bolometric luminosity (L$_{bol}$), and SFR in Table \ref{props}. The dust mass is estimated as is done in \citet{2008MNRAS.388.1595D} summing up the mass contributions from the warm dust in stellar birth clouds, warm dust in the ambient interstellar medium, and the cold dust in the interstellar medium. What is interesting is that not all fits required an AGN component, despite this being an X-ray selected sample. In particular, Bins 3 and 4 achieve good fits without including an AGN component. The increasing contribution of the AGN to the optical emission is clearly seen as the X-ray luminosity of each bin increases.
\par
In addition to performing fits using {\sc Sed3Fit}, we also performed fits using another routine: {\sc xCigale} (the X-ray Code Investigating GALaxy Emission) \citep{2020MNRAS.491..740Y}. For a quick comparison on the results of the two routines, we invite the reader to look through Appendix \ref{app:C}. 
\par
After fitting the Herschel undetected sample, we also fit the SEDs for our cold quasars and full Herschel subsample, which have individual detections with SPIRE, using {\sc Sed3Fit} to allow for a more robust comparison between the two samples. This is in contrast to the method used in \citet{kirkpatrick2020} to calculate SFRs (see Appendix \ref{app:B} for details on how the different fitting methods compare), but it produces consistent results.  

We note that simulations show that in extreme cases, the AGN can account for all of the far-IR/submm heating \citep{mckinney2021}.
Likewise, AGN can account for dust-heating at 70-100\,$\mu$m--a regime attributed to star formation by most of the models in {\sc Sed3Fit} \citep{tadhunter2007,symeonidis2016}. This is mostly a concern for our more luminous bins, and the modeling of our cold quasars. However, our cold quasars and high redshift bins have significantly more dust than low redshift AGN, on which those results are based. Further, in \citet{kirkpatrick2020}, the cold quasars were modeled with a template from \citet{kirkpatrick2012} that has more far-IR AGN heating in it than any other existing, empirical AGN template \citep{lyu2017}. Appendix \ref{app:B} demonstrates that the SFRs of these extreme systems are consistent whether applying {\it Sed3Fit} or the \citep{kirkpatrick2012} template. Similarly, for the stacked bins, the {\sc Sed3Fit} and {\sc X-Cigale} results are largely consistent, despite the inclusion of the X-ray in the {\sc X-Cigale} fitting.

\section{Results}\label{sec:Results}

\subsection{Star formation rate and AGN bolometric luminosity}

\begin{figure*}
	\includegraphics[width=7in]{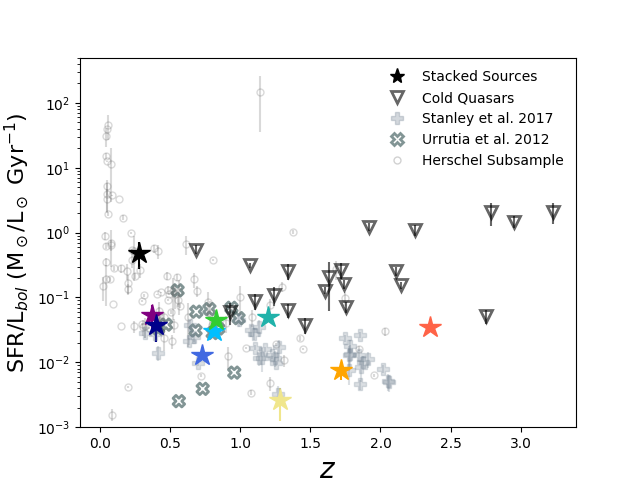}
	\caption{SFR/Lbol is plotted as a function of redshift for the stacked samples (filled stars) compared to the cold quasar sample (open triangles), the \citet{2017MNRAS.472.2221S} data (filled crosses), the Herschel subsample (open circles), and the \citet{2012ApJ...757..125U} data (open X's). The stacked sources are color coded by the bin color in Figure \ref{lxvz}. The stacked sources seem to follow the same downward trend as all other samples except for the cold quasars. The cold quasars have much higher ratios despite their intense luminosities.}
	\label{sfrvlbol}

\end{figure*}

First, we explore whether we see any correlation between the growth of the stellar population and the build-up of black hole mass. Examining the bolometric luminosity of the AGN component will serve as a proxy for black hole growth, as this parameter can be converted to black hole accretion rate assuming a radiative efficiency.  In Figure \ref{sfrvlbol}, we plot the ratio of SFR extracted from the SED fitting and bolometric luminosity as a function of redshift. The bolometric luminosity was determined from $L_{0.5-10\,keV}$ using the bolometric correction in \citet{2007ApJ...654..731H} for both our sample and the \citet{2017MNRAS.472.2221S} sample.
\par
\begin{figure}
	\includegraphics[width=3.7in]{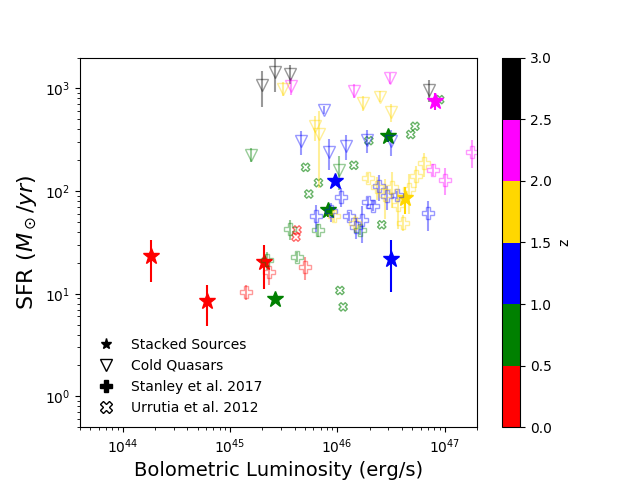}
	\caption{SFR is plotted as a function of bolometric luminosity for the stacked samples (filled stars) compared to the cold quasar sample (open triangles), the \citet{2017MNRAS.472.2221S} data (filled crosses) and the \citet{2012ApJ...757..125U} data (open X's). The symbols are color coded by their mean redshift values. The stacked samples follow a roughly linear trend providing evidence for a {\bf correlation} between host galaxy growth and AGN growth.}
	\label{sfrplot}

\end{figure}

According to Table \ref{props} the SFRs of our stacked sample range from 8 $M_\odot/yr$ to 750 $M_\odot/yr$ keeping in mind that the lower SFRs come from a low redshift and the higher SFRs from higher redshifts. Using these values and the bolometric luminosities of the stacked sources, we see that the ratios follow a flat trend as the redshift increases from 0.5 in Figure \ref{sfrvlbol}. More importantly, this flat trend comes from the linear correlation between the two parameters shown in Figure \ref{sfrplot}, which will be discussed momentarily. 
 
\par
Because different selection methods can be biased towards different physical parameters \citep{2012NewAR..56...93A}, we compare our X-ray selected quasar sample with a sample of optically selected unobscured AGN from SDSS whose SEDs were also constructed via stacking of \textit{Herschel} images \citep{2017MNRAS.472.2221S}. This optically-selected sample is limited to broadline spectroscopic AGN, and 94\% of the sample has WISE detections. Our X-ray selected sample includes a mix of spectroscopic, photometric, broadline, and narrow-line sources, and has a WISE detection rate of 66\%. Nevertheless, we find that the two samples span a consistent range of SFR/$L_{\rm bol}$ with redshift.
\par

More interesting is the fact that the Herschel subsample also spans a similar range in SFR/$L_{bol}$ with redshift as the Herschel-undetected sample. The general increase in SFR/$L_{\rm bol}$ over cosmic time indicates that the black hole growth is slowing relative to stellar growth. This is particularly noticably at $z<0.5$. At $z=1-2$, the star formation rate and bolometric luminosity are correlated, which is echoed in the relatively flat distribution of SFR/$L_{\rm bol}$ at those redshifts. In contrast to the other samples, the cold quasar sample exhibit intense SFRs. On average, they have ratios a factor of 10 higher than our Herschel-undetected sample. 
\par
Red quasars are another population of AGN worthy of comparison. Red quasars are luminous quasars that are moderately obscured ($A_V \approx 1 - 5$) by a cold absorber along their line of sight to the quasar itself, which usually resides in the host galaxy \citep{2012ApJ...757..125U}. \citet{2012ApJ...757..125U} selected 13 of these dust reddened quasars between a redshift of $0.4<z<1.0$ and followed up this sample with the Advanced Camera for Surveys Wide Field Camera on \textit{Hubble Space Telescope} (HST) as well as the MIPS and IRS on board the Spitzer Space Telescope. Red quasars still show broad emission lines in the rest-frame optical and their continuum is still dominated by the quasar rather than the host galaxy. \citet{2012ApJ...757..125U} measured the SFR through multiwavelength SED modeling and determined the bolometric luminosity by integrating over the quasar components. In Figure \ref{sfrvlbol}, we see that the red quasars occupy a broad range of ratios within their narrow redshift range. The differing ratios between the red and cold quasars indicate that these sources are unlikely to be part of the same parent population, which would be the case if geometry was the dominant mechanism driving the differences in SED emission or if they were probing different, narrow phases of the AGN life cycle. 
\par
While Figure \ref{sfrvlbol} can tell us a bit about how the ratio evolves with redshift, it is not obvious whether there is a correlation between the SFR and the bolometric luminosity in our stacked sources. For a direct comparison between the two parameters we plot the SFR as a function of bolometric luminosity in Figure \ref{sfrplot}. This time our points are color coded by their mean redshift. Observing Figure \ref{sfrplot}, not considering the redshift, we see a roughly linear trend between the two parameters for our stacked sample. When we consider the redshift, it can be determined that as the redshift increases, we observe an increase in both the bolometric luminosity and star formation, which is what we expect to see. Similar to Figure \ref{sfrvlbol}, the cold quasars have more intense SFRs on average compared to the rest of the samples. This is an explanation as to why the ratios are much higher in Figure \ref{sfrvlbol}. 

\subsection{Dust Masses \&  X-ray Luminosity}
\begin{figure}
	\includegraphics[width=1.1\linewidth]{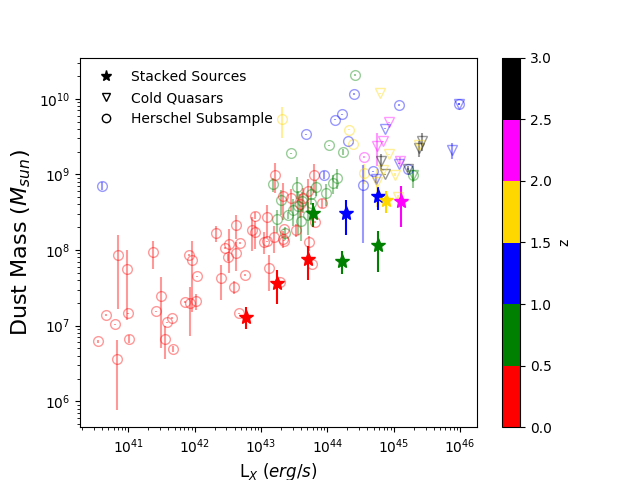}
	\caption{Dust mass as a function of the X-ray luminosity for the stacked samples (filled symbols) compared with the Herschel subsample (open circles) and the cold quasars (open triangles). The samples are color coded by their median redshift. The dust masses of the stacked sources are smaller compared to the Herschel detected samples giving us some information about the ISM of our sample. }
	\label{dusty}
	
\end{figure}

We now examine the dust masses of AGN, which are linked to the interstellar medium and reservoir available for fueling star formation. 
Figure \ref{dusty} shows the relationship between X-ray luminosity and dust mass for our stacked bins, the cold quasars, and the Herschel subsample. 
At a given redshift and X-ray luminosity, the stacked samples exhibit lower dust masses compared to the \textit{Herschel} subsample and cold quasars. This result is expected since Herschel luminosity correlates closely with dust mass. The non-WISE sources likely have even lower dust masses than the stacked bins. Figure \ref{dusty} illustrates that at z$>$0.5 X-ray luminosity is not correlated with dust mass, as the stacked and Herschel samples span quite different ranges. Dust mass typically traces the cold gas reservoir \citep{groves2015,scoville2016}. It seems, then, that the global amount of galactic gas available for fueling stellar growth does not have a direct, predictable impact on the central gas fueling black hole growth.

\section{Discussion}\label{sec:Discussion}
The evolution of a galaxy is commonly understood through the main sequence of galaxies, which involves a tight relationship between SFR and stellar mass that is exhibited by most star-forming galaxies \citep{2007A&A...468...33E,2007ApJ...660L..43N}.
Galaxies are categorized in terms of two different modes of star formation: a ``normal" mode such as a galaxy on the main sequence, and a starburst mode (short-lived, highly star-forming galaxies above the main sequence), with the former being most associated with local disk galaxies and the latter with ultra/luminous infrared galaxies and/or high-$z$ sub-millimeter galaxies \citep{2012ApJ...757...23K,2018ApJ...865..103C,kirkpatrick2019}. Normal star-forming galaxies have gas depletion time scales ($\tau_{\rm dep}$) that are greater than that of starburst galaxies. When the gas depletion reaches the point to where there is little to no gas remaining, star formation slows to a halt. This process is known as quenching. As the star formation in a galaxy shuts down, it leaves the main sequence. Because our sample is undetected by Herschel, it comes into question whether our stacked sources are on the main sequence or not. If AGN are indeed linked with quenching, we may expect to find quasars below the main sequence.

\par
\begin{figure}
	\includegraphics[width=3.5in]{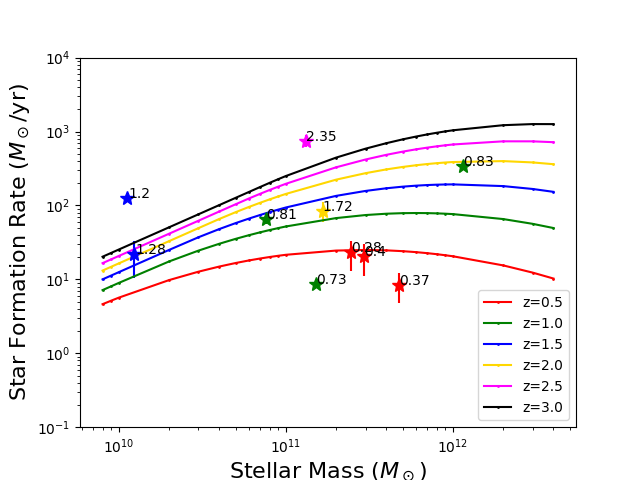}
        \caption{Relationship between the SFR and stellar masses of our stacked sources. Each of the sources have their median redshift attached for comparison. Each colored line on the plot represents the galaxy main sequence line at the given redshift using the main sequence equation \citep{2015A&A...575A..74S}. The symbols are color coded to be the same color as the upper main sequence line if the mean redshift falls between successive lines. Some sources are offset from their respective main sequence line.}
    \label{mainseq}
\end{figure}

In Figure \ref{mainseq}, we show the location of our stacked galaxies on a plot of SFR vs. $M_*$, specifying the mean redshift of the stack next to each data point. We show for comparison the main sequence using the star-forming main sequence fits in \citet{2015A&A...575A..74S}, which parametrized the evolution of the main sequence with redshift. The different colored lines represent the redshift range spanned by our sources and the symbols are color coded to the rounded up main sequence line. Any point that has a mean redshift less than 0.5 will be colored purple and so on. Figure \ref{mainseq} illustrates where the stacked sources fall according to the different main sequence lines. However, just on the basis of Figure \ref{mainseq}, it is not obvious where they lie compared to the main sequence. We observe offsets from the main sequence lines for some of the sources, which makes it difficult to determine if the sources are in the main sequence region. However, we note that the redshifts for the main sequence lines are different than the redshifts for the sources.
\par
\begin{figure}[ht]
	\includegraphics[width=3.7in]{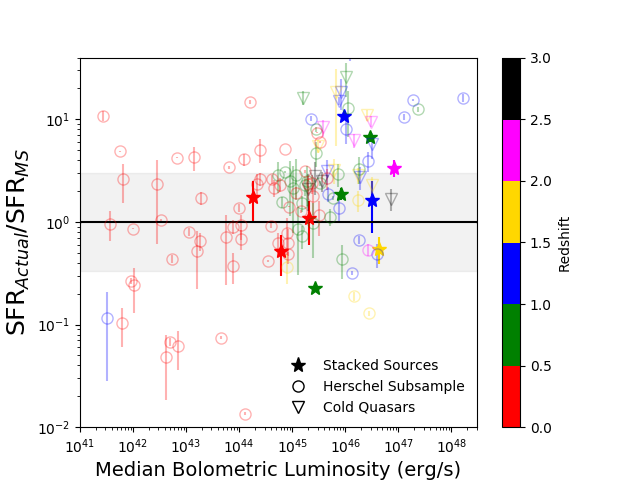}
        \caption{The main sequence ratio for our stacked sources (filled stars) as a function of bolometric luminosity compared to the Herschel subsample (open circles) and cold quasars (open triangles). The grey region represents the main sequence region and ranges from $0.33-3.0$. The sources below a ratio of 0.33 are considered to be heavily quenched. The majority of the stacked sources lie on or above the main sequence, which is similar to the Herschel subsample. This suggests that X-ray bright AGN show no signs of quenching.}
    \label{resid}
\end{figure}

We determine the ratio of the source SFR and main sequence SFR our sources possess by calculating the SFR they would have if 
they were on the main sequence using the same equation in \citet{2015A&A...575A..74S} that was used for Figure \ref{mainseq}. We then take a ratio of the expected SFR and the value derived from our SED fits. Anything above the main 
sequence (ratio $>$ 3) is considered to be a starburst, and anything below (ratio $<$ 0.33) is considered to be 
quenching. Comparing our sample with the cold quasar sample in Figure \ref{resid}, we see that 60\% of the cold quasars lie above the shaded region and are therefore considered to be starbursts, which is expected given the extreme SFRs highlighted in \citet{kirkpatrick2020}. Furthermore, six of our stacked samples are considered to be on the main sequence (gray shaded region), three are considered to be starburst galaxies, and only one is considered to be quenching. An interesting note from this plot is that the ratio is not highly dependent on either stellar mass or AGN bolometric luminosity, but may have a correlation with redshift. As the redshift increases we see an average increase in the main sequence residual. Figure \ref{resid} suggests that it is difficult to conclude that quenching is present in X-ray selected AGN. Nor does X-ray selection alone predict that a galaxy will be found in a certain location on the main sequence, as all of our sources span a wide range. This is in contrast to what is expected from the major merger scenario which is that there is a specific timeline for the life cycle of an AGN and IR faint AGN  should have declining star formation \citep{1988ApJ...328L..35S,2006ApJ...652..864H,2017MNRAS.468.1273R,2010Sci...328..600T}.

\begin{figure}
    \centering
    \includegraphics[width=1.1\linewidth]{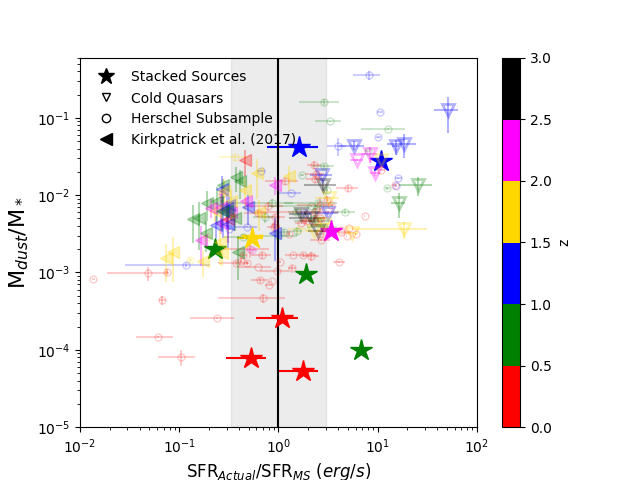}
    \caption{The Dust-to-Stellar Mass ratio plotted as a function of Main sequence distance compared to the Herschel subsample (open circles), cold quasars (open triangles) and dusty star forming galaxies from \citet{2017ApJ...843...71K} (filled triangles). The grey region represents the main sequence region (0.33 - 3.0). We observe the stacked sources to have mass ratios consistent with SFGs and a noticeable trend with redshift. We expect the mass ratio to increase with redshift due to the increase in SFR for high redshift sources.}
    \label{dust2stell}
\end{figure}
\par
Next, we assess the relationship between the dust mass and stellar mass of our sources and their consistency with main sequence SFGs. This mass ratio is roughly a proxy for the gas-to-stellar mass ratio. We plot this ratio as a function of the main sequence distance (Figure \ref{resid}) in Figure \ref{dust2stell}. Comparing to the dusty star forming galaxy sample from \citet{2017ApJ...843...71K}, the mass ratios are consistent for the stacked sources. Furthermore, we observe a redshift trend with the mass ratio. This is to be expected since we see an increase in SFR for the stacked sources at higher redshift.

\par
\begin{figure}
    \centering
	\includegraphics[width=1.1\linewidth]{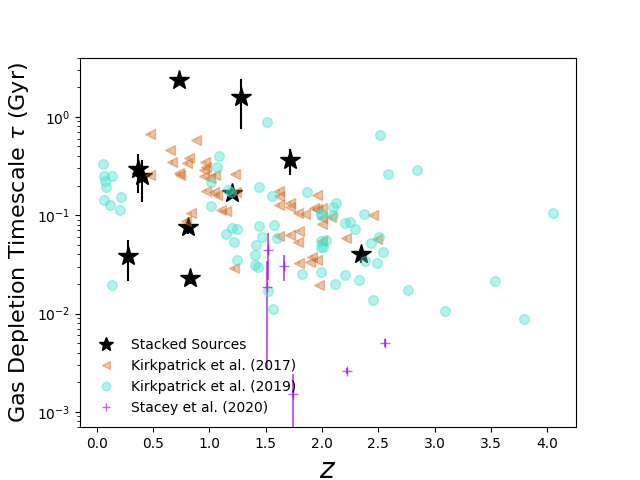}
    \caption{Gas depletion timescale as a function of redshift for our stacked sources (black symbols) compared to dusty star forming galaxies from \citet{2017ApJ...843...71K} (brown triangles), IR selected AGN from \citet{kirkpatrick2019} (turquoise circles), and gravitationally lensed quasars from \citet{2020arXiv200901277S} (purple plus signs). on average our stacked SEDs have similar $\tau$ as IR-selected AGN and star forming galaxies. For some of the data points, the errors are within the size of the points.}
    \label{sfe}
\end{figure}

We also investigate the gas depletion timescale ($\tau_{\rm dep}= M_{\rm gas}/SFR$) of our sources to see how the Herschel undetected sources compare with dustier galaxies. We plot this quantity as a function of redshift in Figure \ref{sfe}, where we convert M$_{dust}$ to M$_{gas}$ using the relation

\begin{equation}
    \log M_{\rm mol.\,gas} = \log M_{\rm dust} + 1.83
\end{equation}

from a study on Stripe82 galaxies at z $<$ 0.2 \citep{2018MNRAS.478.1442B}. We note that the original relation includes a very slight dependence on metallicity, which we have removed, as we cannot measure this very accrutely for our galaxies. In their given metallicy range (based on stellar mass estimates), including a metallicity dependence would change the molecular gas mass by at most 9\%. 
 
$\tau_{\rm dep}$ is correlated with redshift, and we find no significant correlation with AGN bolometric luminosity, as also seen in \citep{kirkpatrick2019}. We compare to a sample of IR-selected dusty star forming galaxies from \citet{2017ApJ...843...71K}, the sample of IR-selected AGN from \citet{kirkpatrick2019}, and 7 gravitationally lensed quasars which are bright in the submm from \citet{2020arXiv200901277S}. We see that on average our stacked sources have similar $\tau_{\rm dep}$ compared to the \citet{2017ApJ...843...71K} and \citet{kirkpatrick2019} samples. The lensed quasars have faster $\tau_{\rm dep}$, due to their much more extreme SFRs (SFR$\sim1000\,M_\odot$/yr) compared with the other samples. 

Our stacked galaxies span a range of $\tau_{\rm dep}$, with most sources have $\tau_{\rm dep}\sim100\,$Myr. We also have two bins with very long $\tau_{\rm dep}$ of over 1 Gyr. 
 
Because of the similarities between the stacked sources and SFGs, it is still difficult to conclude that quenching is present, as quenching typically follows from high star formation efficiencies and short gas depletion timescales.  
This could indicate that the AGN itself is not responsible for shutting down star formation. We note here that timescales may influence this conclusion. X-ray emission traces the instantaneous AGN luminosity, while IR-based SFRs are averaged over the past 100\,Myr. Quenching may be about to set in in these galaxies, but we would not be able to tell due to the difference in timescales traced. This is a perennial struggle in AGN studies.
\par
We find that Herschel-undetected, X-ray luminous AGN are still forming stars at rate comparable to the Herschel-detected AGN. However, their gas masses are on average lower than Herschel-detected sources. In the Herschel-undetected sources, then, the star formation must be less obscured than in the Herschel-detected sources. The far-IR emission is independent of the emission of the AGN, traced through $L_X$ and $L_{\rm bol}$ and obscuration (as our sample contained a mix of obscured and unobscured sources while \citep{2017MNRAS.472.2221S} contained only unobscured sources). Far-IR emission depends only on the amount of obscuring dust in the host galaxy itself, surrounding the sites of star formation. Whatever fueling mechanism funnels gas down to the central parsec around the AGN has apparently little relation the amount of gas in the host galaxy.

\section{Conclusions}\label{sec:conclusions}
We present a sample of X-ray bright, Herschel-undetected AGN from the Stripe 82X field. We binned our sample by their X-ray luminosity and redshift. Our sources observed have $L_X > 10^{42} erg/s$ and a redshift range of $z \approx 0-3$.  We performed IR stacking using \textit{Herschel/SPIRE} maps to obtain a mean IR detection in all three Herschel bands. We created stacked SEDs from all of the other photometric filters available for this sample, ranging from the UV to the mid-IR. We took those 10 composite photometry sets and ran them through the SED fitting code {\sc Sed3Fit} and used the fit results to extract estimates of SFR, stellar mass, dust mass, and IR luminosity. We then compared our findings with various AGN samples. Our findings include
\begin{enumerate}
    \item On average, the stacked sources span a similar range of SFR/L$_{bol}$ not only as the optically selected sample but also as the \textit{Herschel} subsample.  
    \item According to Figure \ref{sfrvlbol}, we observe a flat trend between the stellar population and build up of black hole mass at z$>$0.5.
    \item At z$>$0.5, X-ray luminosity is not correlated with dust mass suggesting that the amount of gas available for stellar growth does not have a predictable impact on the central gas fueling black hole growth. 
    \item Despite a smaller dust mass than the \textit{Herschel} subsample, the majority of the stacked sources lie on or above the main sequence, which is similar to the Herschel subsample. X-ray selection alone does not predict the location on the main sequence a galaxy will be found.
    \item With similar $\tau_{dep}$ values as dusty star forming galaxies, it is difficult to conclude the presence of quenching in X-ray selected AGN. 
\end{enumerate}

When it comes to understanding the far-IR properties of Herschel undetected, X-ray bright AGN, there is more work that needs to be done which requires future sensitive IR space telescopes capable of measuring the ISM in individual galaxies. Furthermore, with lower dust masses, but no signs of quenching, we must begin to explore the properties of the ISM in more detail.
\newline
\par

This research is based upon work supported by NASA under award No.
80NSSC18K0418 to Yale University and by the National Science Foundation under Grant No. AST-1715512.
\par
Based in part on observations made with the NASA/DLR Stratospheric Observatory for Infrared Astronomy (SOFIA). SOFIA is jointly
operated by the Universities Space Research Association,
Inc. (USRA), under NASA contract NNA17BF53C, and
the Deutsches SOFIA Institut (DSI) under DLR contract
50 OK 0901 to the University of Stuttgart. 
\par
ET acknowledges support from FONDECYT Regular 1190818 and 1200495, ANID grants CATA-Basal AFB-170002, ACE210002, and FB210003, and Millennium Nucleus NCN19\_058.
\par
This research made use of Astropy, a community developed core Python package for Astronomy \citep{2013A&A...558A..33A,2018AJ....156..123A}. This research made use of the iPython environment \citep{2007CSE.....9c..21P} and the Python packages SciPy \citep{2020zndo....595738V}, NumPy \citep{2011CSE....13b..22V}, and
Matplotlib \citep{2007CSE.....9...90H}.

\clearpage
\appendix

\section{Appendix A}\label{app:A}

\begin{deluxetable}{c ccc ccc ccc c}[h]
\tabletypesize{\footnotesize}
\tablecolumns{11}
\tablewidth{0pt}
\tablecaption{Mean Photometry (mJy)}
\tablehead{
\colhead{Filter} & \colhead{Bin 1} & \colhead{Bin 2} & \colhead{Bin 3} & \colhead{Bin 4} & \colhead{Bin 5} & \colhead{Bin 6} & \colhead{Bin 7}  & \colhead{Bin 8} & \colhead{Bin 9} & \colhead{Bin 10}}
\startdata
u   & 0.205     & 0.699     & 0.030     & 0.007     & 0.014     & 0.010     & 0.129     & 0.028     & 0.024     & 0.020     \\
    & (0.154)   & (0.671)   & (0.008)   & (0.002)   & (0.002)   & (0.001)   & (0.071)   & (0.002)   & (0.003)   & (0.003)   \\
g   & 0.665     & 0.877     & 0.066     & 0.012     & 0.021     & 0.011     & 0.188     & 0.030     & 0.028     & 0.036     \\
    & (0.466)   & (0.821)   & (0.018)   & (0.003)   & (0.003)   & (0.001)   & (0.105)   & (0.002)   & (0.003)   & (0.008)   \\
r   & 1.17      & 1.35     & 0.158     & 0.023     & 0.026     & 0.015     & 0.286     & 0.040     & 0.030     & 0.044     \\
    & (0.82)    & (1.21)   & (0.055)   & (0.005)   & (0.004)   & (0.001)   & (0.187)   & (0.003)   & (0.003)   & (0.012)   \\
i   & 1.78      & 1.50     & 0.365     & 0.064     & 0.037     & 0.016     & 0.509     & 0.042     & 0.039     & 0.050     \\
    & (1.06)    & (1.22)   & (0.175)   & (0.021)   & (0.008)   & (0.001)   & (0.368)   & (0.003)   & (0.004)   & (0.015)   \\
z   & 1.80      & 2.30      & 0.622     & 0.134     & 0.053     & 0.020     & 0.737     & 0.043     & 0.040     & 0.059     \\
    & (0.91)    & (1.73)   & (0.336)   & (0.055)   & (0.012)   & (0.001)   & (0.538)   & (0.003)   & (0.004)   & (0.016)   \\
\hline
J   & 2.36      & 2.53       & 0.853     & 0.238     & 0.077     & 0.027     & 1.40      & 0.050     & 0.047     & 0.063     \\
    & (1.04)    & (1.73)   & (0.501)   & (0.107)   & (0.020)   & (0.001)   & (1.09)    & (0.003)   & (0.005)   & (0.016)   \\
H   & 1.84      & 2.78       & 0.838     & 0.247     & 0.088     & 0.035     & 0.943     & 0.063     & 0.054     & 0.067     \\
    & (0.72)    & (1.93)   & (0.444)   & (0.097)   & (0.021)   & (0.001)   & (0.666)   & (0.004)   & (0.006)   & (0.015)   \\
K   & 1.97      & 3.53       & 0.909     & 0.253     & 0.104     & 0.049     & 1.30      & 0.070     & 0.059     & 0.077     \\
    & (0.78)    & (2.65)    & (0.446)   & (0.095)   & (0.018)   & (0.002)   & (0.97)    & (0.004)   & (0.006)   & (0.012)   \\
\hline
W1  & 1.84      & 5.44       & 0.696     & 0.226     & 0.150     & 0.084     & 5.56      & 0.121     & 0.087     & 0.077     \\
    & (0.85)    & (4.73)   & (0.253)   & (0.067)   & (0.012)   & (0.004)   & (4.90)    & (0.006)   & (0.007)   & (0.007)   \\
W2  & 1.09      & 3.13       & 0.657     & 0.191     & 0.175     & 0.116     & 3.21      & 0.197     & 0.141     & 0.106     \\
    & (0.46)    & (2.61)   & (0.173)   & (0.046)   & (0.014)   & (0.007)   & (2.53)    & (0.012)   & (0.011)   & (0.008)   \\
W3  & 0.715     & 1.16      & 1.14      & 0.514     & 0.560     & 0.465     & 1.85      & 0.613     & 0.590     & 0.551     \\
    & (0.096)   & (0.48)   & (0.15)    & (0.027)   & (0.027)   & (0.016)   & (0.55)    & (0.032)   & (0.036)   & (0.034)   \\
W4  & 2.86      & 3.29        & 3.99      & 2.84      & 2.81      & 2.80      & 4.80      & 2.92      &  2.91      & 2.79      \\
    & (0.10)    & (0.25)    & (0.36)    & (0.07)    & (0.07)    & (0.06)    & (0.62)    & (0.06)    & (0.08)    & (0.08) \\
\enddata
\tablecomments{Uncertainties on the fluxes are derived from bootstrapping and are listed in parenthesis. The filters u, g, i, r, and z are taken from SDSS. The filters J, H, and K are all taken from the VISTA VHS survey. The filters W1, W2, W3, and W4 are from WISE.}
\label{photometry}
\end{deluxetable}

\begin{deluxetable}{c ccc ccc ccc c}
\tabletypesize{\footnotesize}
\tablecolumns{11}
\tablewidth{0pt}
\tablecaption{Mean Photometry (No mock) (mJy)}
\tablehead{
\colhead{Filter} & \colhead{Bin 1} & \colhead{Bin 2} & \colhead{Bin 3} & \colhead{Bin 4} & \colhead{Bin 5} & \colhead{Bin 6} & \colhead{Bin 7}  & \colhead{Bin 8} & \colhead{Bin 9} & \colhead{Bin 10}}
\startdata
u det.[$\Delta$F]   & 83     &   140    & 66  & 203     & 152     & 169     & 55     & 180     & 202     & 126     \\
    & [0.00]   & [0.251]   & [0.00]   & [0.001]   & [0.006]   & [0.00]   & [0.061]   & [0.021]   & [0.01]   & [0.005]   \\
g det.[$\Delta$F]   & 83     & 143     & 66     & 214     & 156     & 169     & 55     & 180     & 201     & 132     \\
    & [0.00]   & [0.024]   & [0.00]   & [0.00]   & [0.004]   & [0.00]   & [0.076]   & [0.005]   & [0.008]   & [0.011]   \\
r det.[$\Delta$F]   & 83      & 142     & 66     & 214     & 157     & 169     & 55     & 180     & 205     & 131     \\
    & [0.00]   & [0.035]   & [0.00]   & [0.00]   & [0.006]   & [0.00]   & [0.121]   & [0.001]   & [0.007]   & [0.014]   \\
i det.[$\Delta$F]   & 82      & 141     & 66     & 214     & 158     & 169     & 55     & 180     & 206     & 133     \\
    & [0.005]   & [0.029]   & [0.00]   & [0.00]   & [0.007]   & [0.00]   & [0.262]   & [0.006]   & [0.005]   & [0.012]   \\
z det.[$\Delta$F]   & 83      & 143      & 66     & 214     & 161     & 169     & 57     & 182     & 208      & 132     \\
    & [0.00]   & [0.009]   & [0.00]   & [0.00]   & [0.003]   & [0.00]   & [0.025]   & [0.007]   & [0.001]   & [0.009]   \\
\hline
J det.[$\Delta$F]    & 83      & 142       & 66     & 202     & 159     & 162     & 58      & 176     & 179     & 112     \\
    & [0.00]   & [0.012]   & [0.00]   & [0.121]   & [0.025]   & [0.001]   & [0.00]   & [0.01]   & [0.011]   & [0.02]   \\
H det.[$\Delta$F]    & 83      & 143       & 66     & 197     & 153     & 162     & 56     & 176     & 173     & 112     \\
    & [0.00]   & [0.008]   & [0.00]   & [0.086]   & [0.028]   & [0.001]   & [0.061]   & [0.008]   & [0.014]   & [0.021]   \\
K det.[$\Delta$F]    & 83      & 141       & 66     & 203     & 158     & 154     & 58      & 171     & 163     & 109     \\
    & [0.00]   & [0.011]   & [0.00]   & [0.093]   & [0.026]   & [0.001]   & [0.00]   & [0.006]   & [0.012]   & [0.023]   \\
\hline
W1 det.[$\Delta$F]   & 83      & 144       & 66     & 214     & 176     & 169     & 58      & 191     & 223     & 135     \\
    & [0.00]    & [0.00]   & [0.00]   & [0.00]   & [0.00]   & [0.00]   & [0.00]    & [0.00]   & [0.00]   & [0.00]   \\
W2 det.[$\Delta$F]   & 83      & 144       & 66     & 214     & 176     & 169     & 58      & 191     & 223     & 135     \\
    & [0.00]    & [0.00]   & [0.00]   & [0.00]   & [0.00]   & [0.00]   & [0.00]    & [0.00]   & [0.00]   & [0.00]   \\
W3 det.[$\Delta$F]   & 83      & 144       & 66     & 214     & 176     & 169     & 58      & 191     & 223     & 135     \\
    & [0.00]    & [0.00]   & [0.00]   & [0.00]   & [0.00]   & [0.00]   & [0.00]    & [0.00]   & [0.00]   & [0.00]   \\
W4 det.[$\Delta$F]   & 83      & 144       & 66     & 214     & 176     & 169     & 58      & 191     & 223     & 135     \\
    & [0.00]    & [0.00]   & [0.00]   & [0.00]   & [0.00]   & [0.00]   & [0.00]    & [0.00]   & [0.00]   & [0.00]   \\
\enddata
\tablecomments{Because we removed any sources without a W1 detection, all sources in each bin were detected at W1. Furthermore, all sources that had a W1 detection were also detected in the other WISE bands. Therefore, mock photometry was not needed for the WISE bands.}
\label{nomock}
\end{deluxetable}

\newpage

\section{Appendix B}\label{app:B}
In \citet{kirkpatrick2020}, the SFRs of the Herschel sample were determined through a far-IR decomposition fitting a single stellar and a single AGN template, where the AGN template allows for a substantial amount of far-IR heating \citep{kirkpatrick2012}. The bolometric luminosities were determined in the same manner as in this paper, by applying a correction to the X-ray luminosity.

\begin{figure*}[h]
    \centering
    \includegraphics{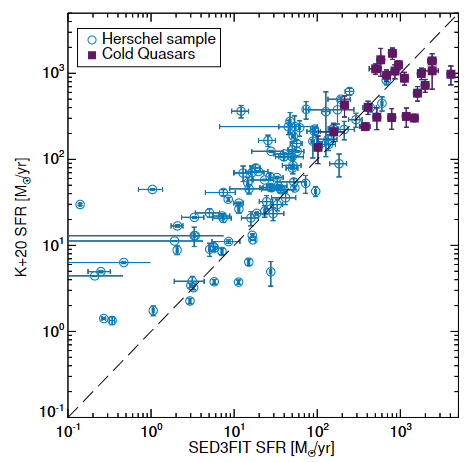}
    \caption{The star formation rates of the Herschel subsample (Cold Quasars are shown as filled purple squares). The x-axis is the SFRs determined with the {\sc Sed3Fit} code and the y-axis are the SFRs measured with a simpler IR decomposition discussed in \citet{kirkpatrick2020}. The SFRs are consistent, with the \citet{kirkpatrick2020} rates being slightly higher.} 
    \label{fig:compare}
\end{figure*}

Figure \ref{fig:compare} shows \citet{kirkpatrick2020} SFRs in comparison with the {\sc Sed3Fit} values used in this paper. There is a one-to-one correlation, although the \citet{kirkpatrick2020} values are generally higher. The median $SFR^{\rm K20}/SFR^{\rm SED3FIT}=1.7$. Both methods take into account the IR AGN emission before calculating the SFR. The cold quasars have no such offset. Using the \citep{kirkpatrick2020} values would have the net effect of raising the Herschel sample in Figure \ref{resid}.

\clearpage

\section{Appendix C}\label{app:C}

In this paper, we used the fitting routine {\sc Sed3Fit} to create best fit models to our stacked SEDs and extract the parameters needed for this study. To test the robustness of the fits, we also created best fit models using another fitting routine {\sc xCigale}. 
\par
{\sc xCigale} is an updated version of {\sc Cigale} which fits SEDs of extragalactic sources \citep{2005MNRAS.360.1413B,2009A&A...507.1793N,2011ApJ...740...22S,2019A&A...622A.103B}. It employs physical AGN and galaxy models, and allows flexible combination between them. However, {\sc Cigale} is not able to model X-ray fluxes, which provides a unique view of AGN \citep{2020MNRAS.491..740Y}. Therefore, \citet{2020MNRAS.491..740Y} builds upon {\sc Cigale} and allows it to model X-ray fluxes and improves the fitting from UV to IR and calls the improved code {\sc xCigale}. It utilizes the $\alpha_{ox}-L_{2500 A}$ relation, which is derived directly from observations of unobscured AGN in order to connect the X-rays to the UV \citep{2006AJ....131.2826S,2007ApJ...665.1004J,2017A&A...602A..79L}. This allows the X-ray data to be fit simultaneously with the other multi-wavelength data. The code allows the input of model parameters. It then realizes the model SED and convolves the model SED with the filters to derive model fluxes. Then comparing the models to the observed fluxes, it computes the likelihood for each model. The code supports both a maximum likelihood analysis and Bayesian-like analysis. In the maximum likelihood analyses, it picks out the model with the largest likelihood and computes physical properties like SFR and stellar mass. Bayesian-like analyses involve computing a probability distribution function and deriving the mean and standard deviation of the physical properties from that. 
\par
\begin{figure*}[ht]
    \centering
    \includegraphics[width=0.33\textwidth]{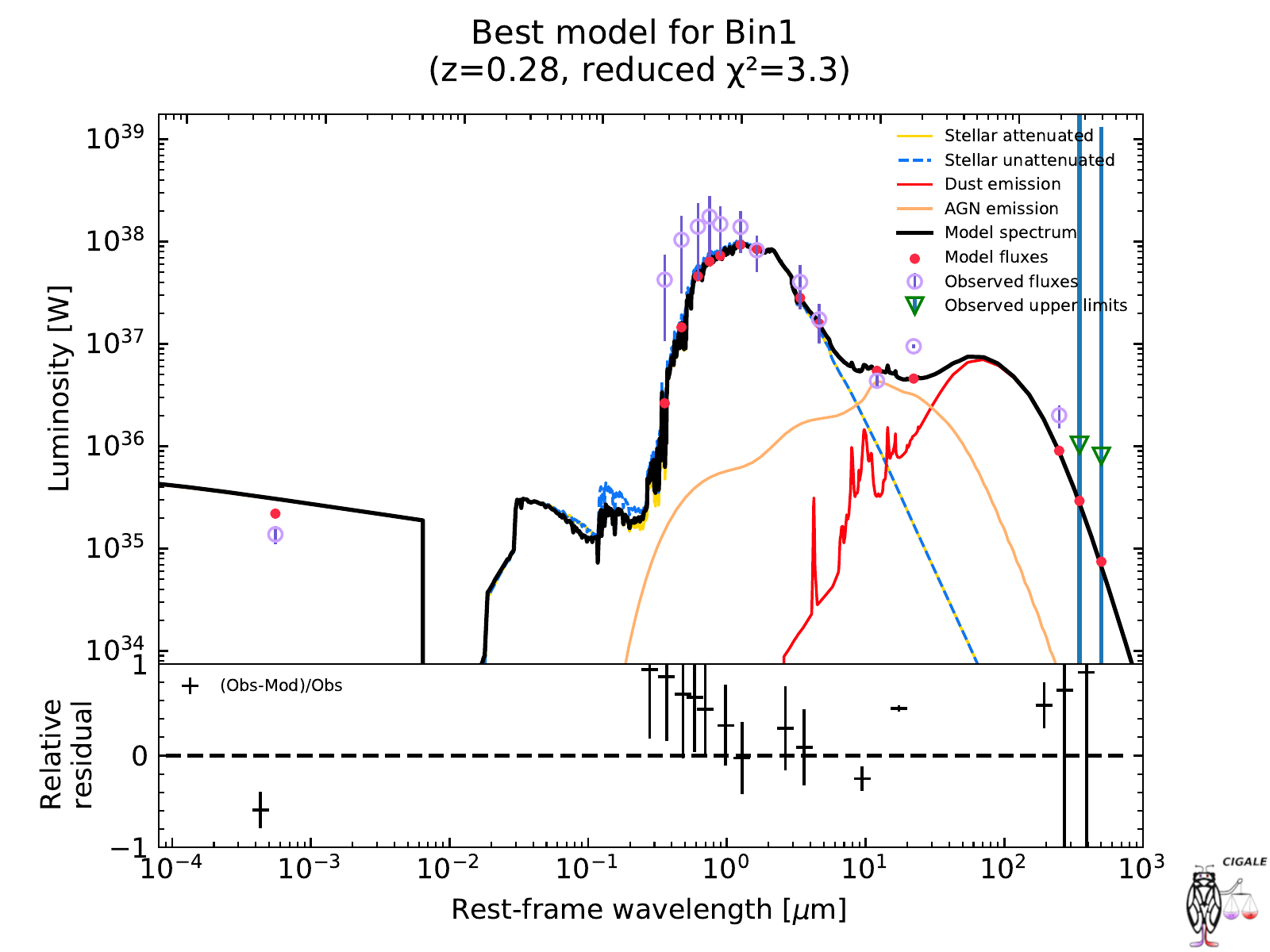}
    \includegraphics[width=0.33\textwidth]{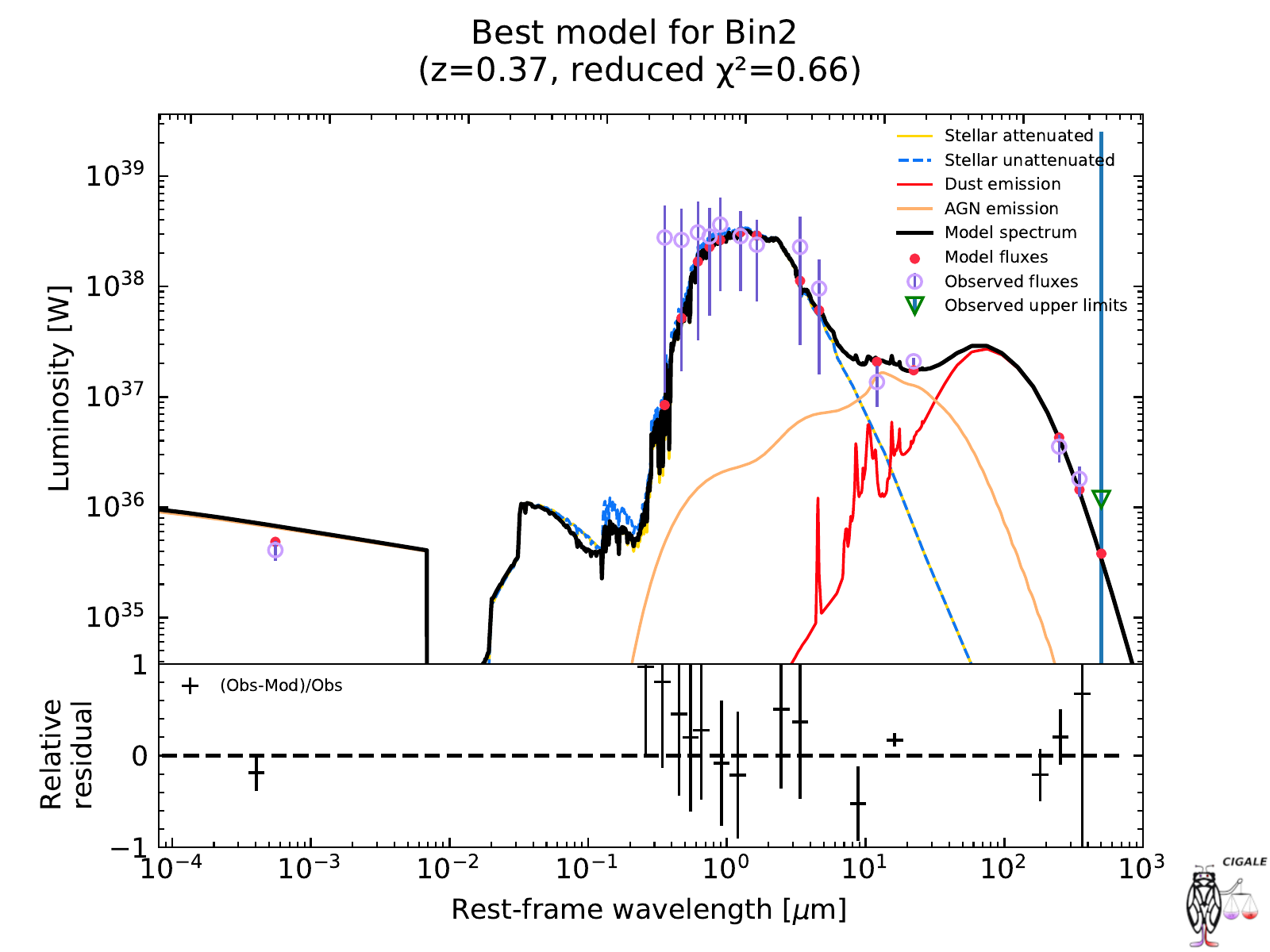}
    \includegraphics[width=0.33\textwidth]{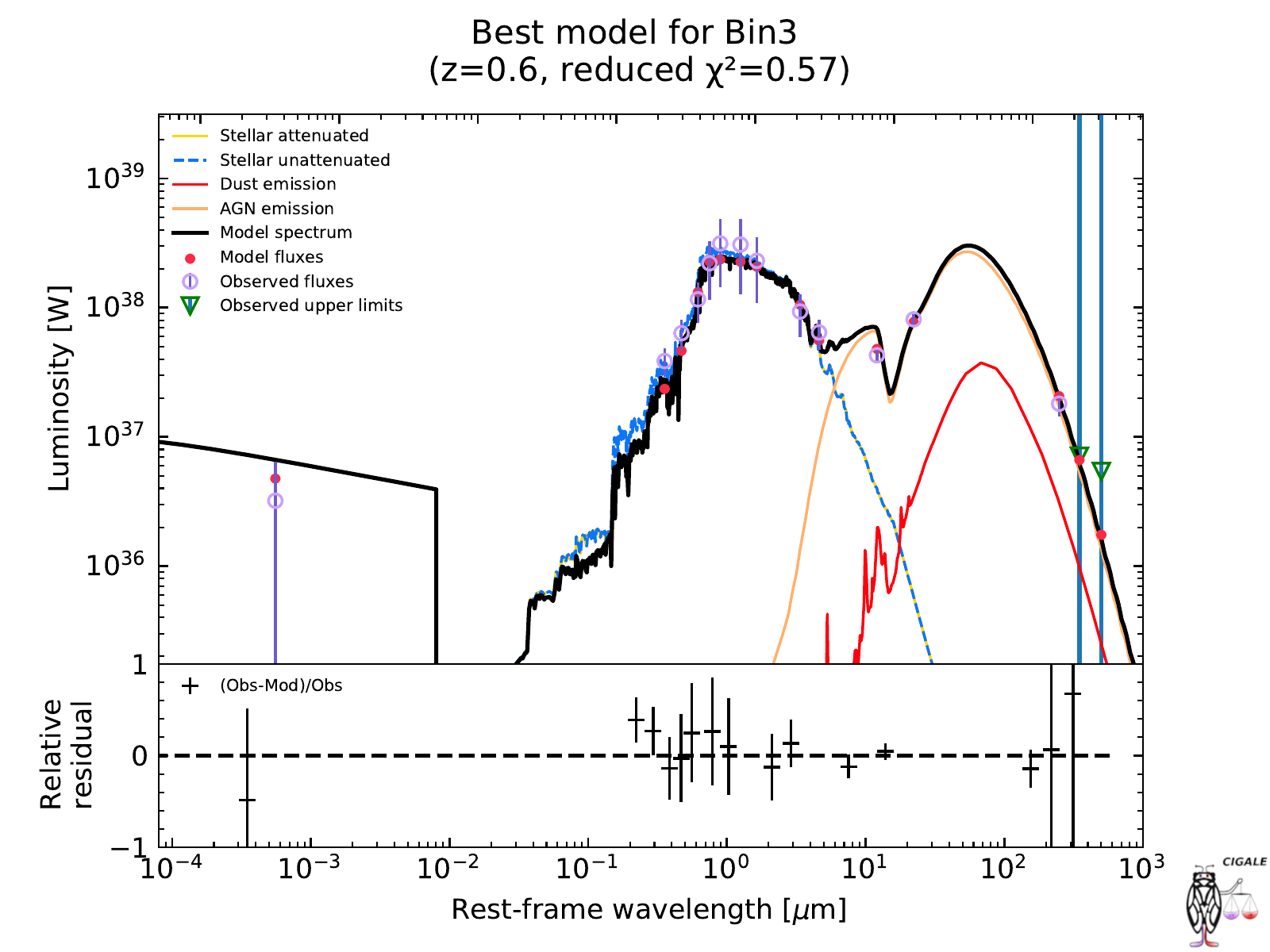}
    \includegraphics[width=0.33\textwidth]{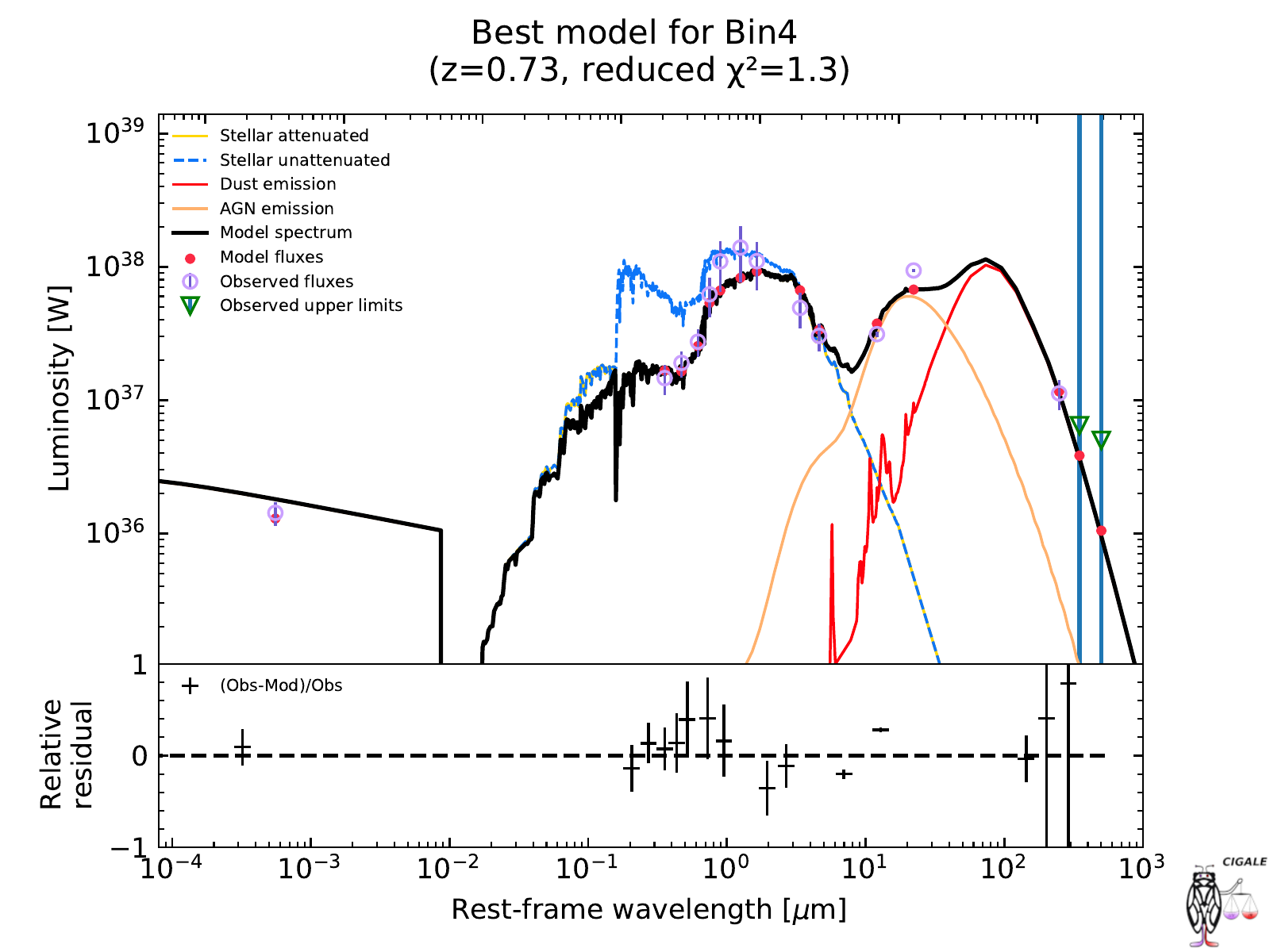}
    \includegraphics[width=0.33\textwidth]{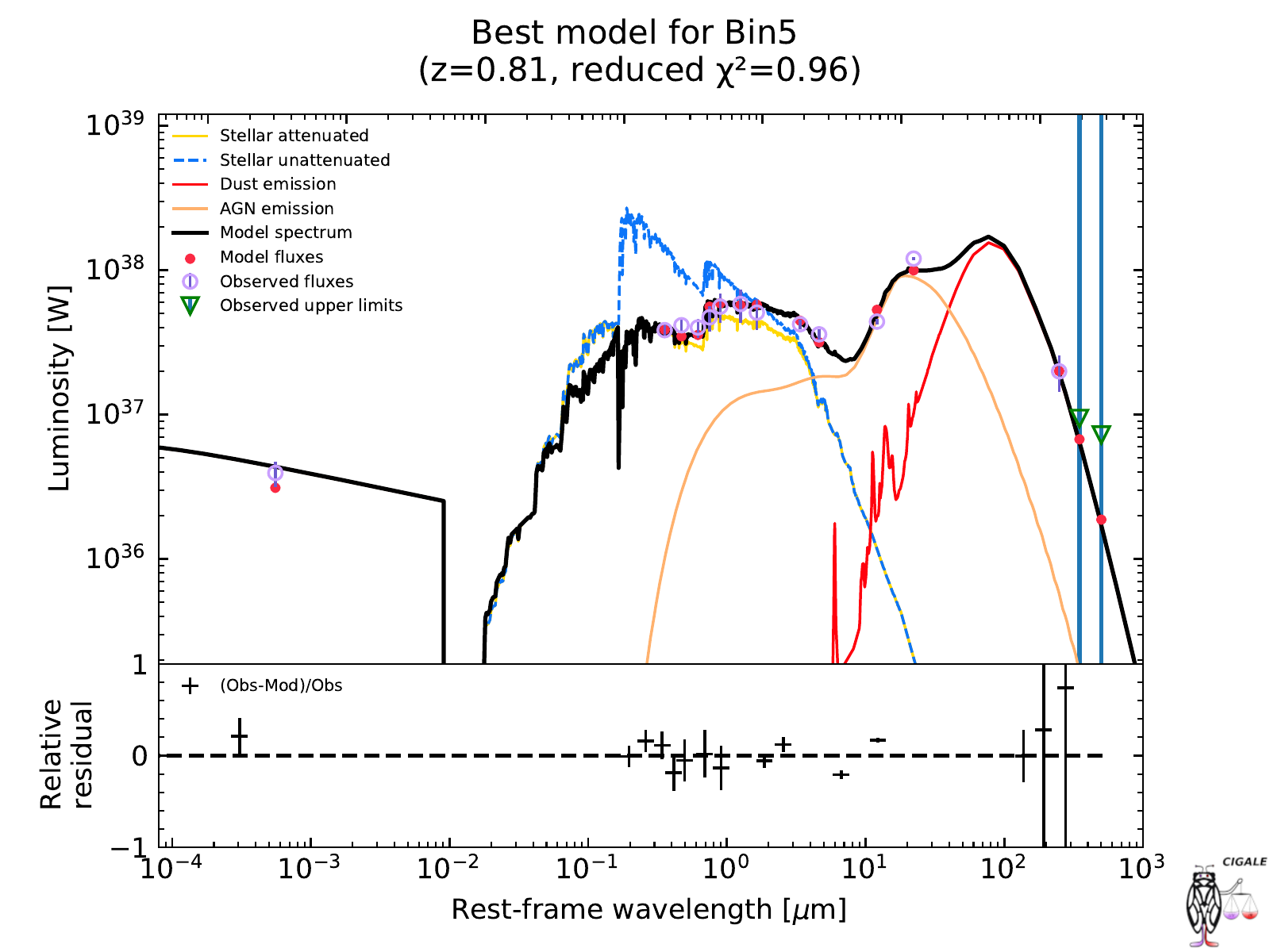}
    \includegraphics[width=0.33\textwidth]{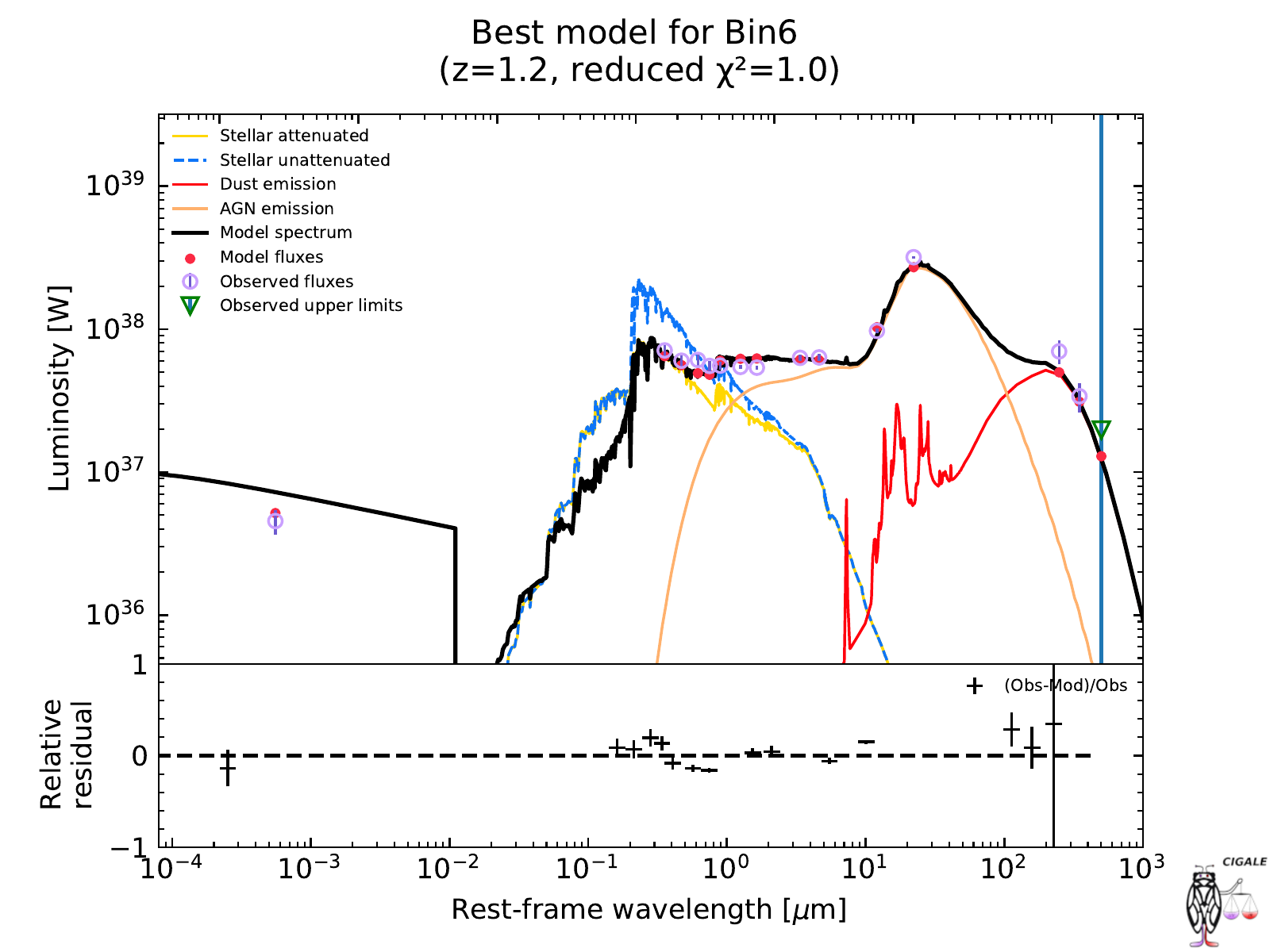}
    \includegraphics[width=0.33\textwidth]{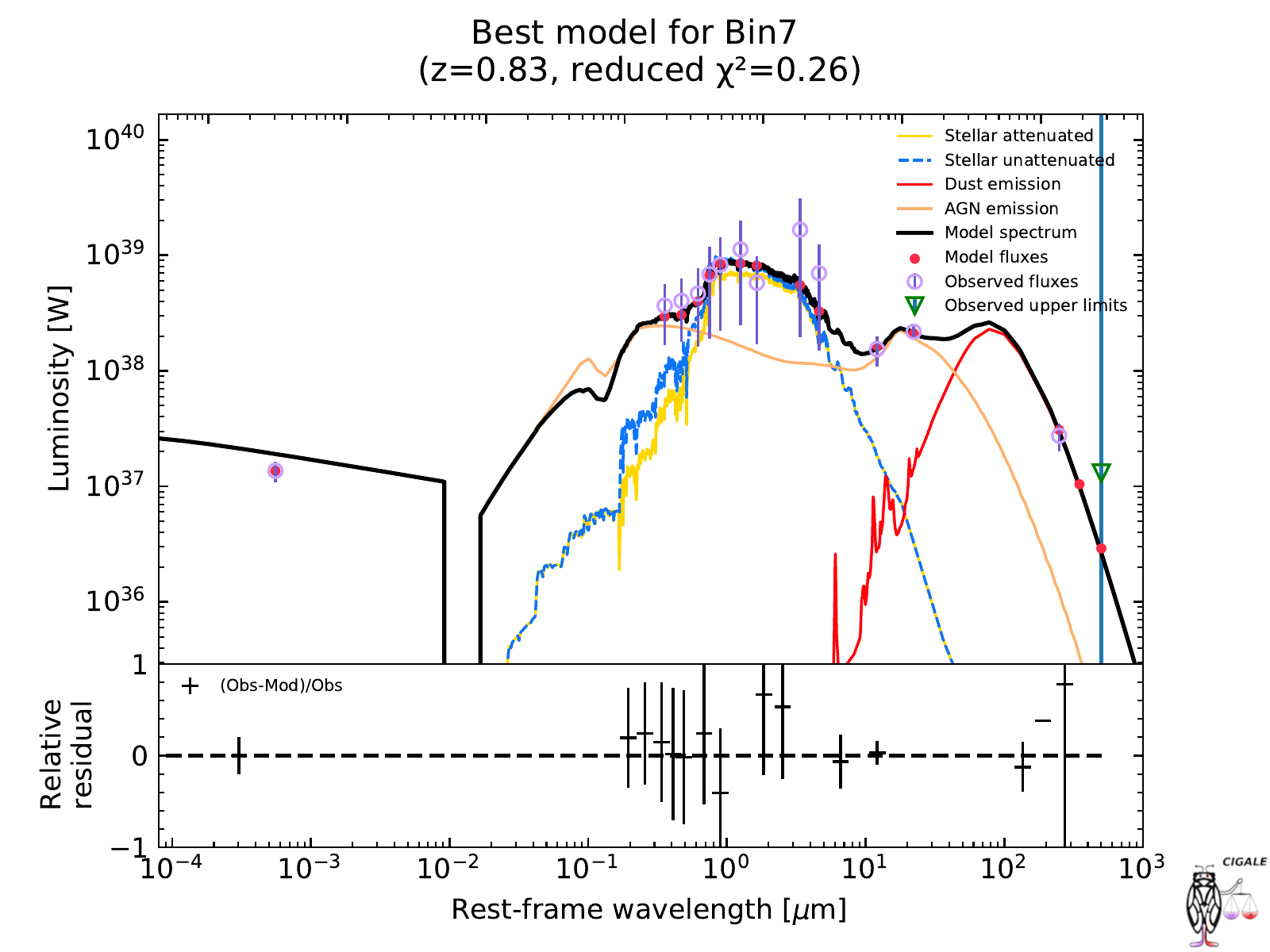}
    \includegraphics[width=0.33\textwidth]{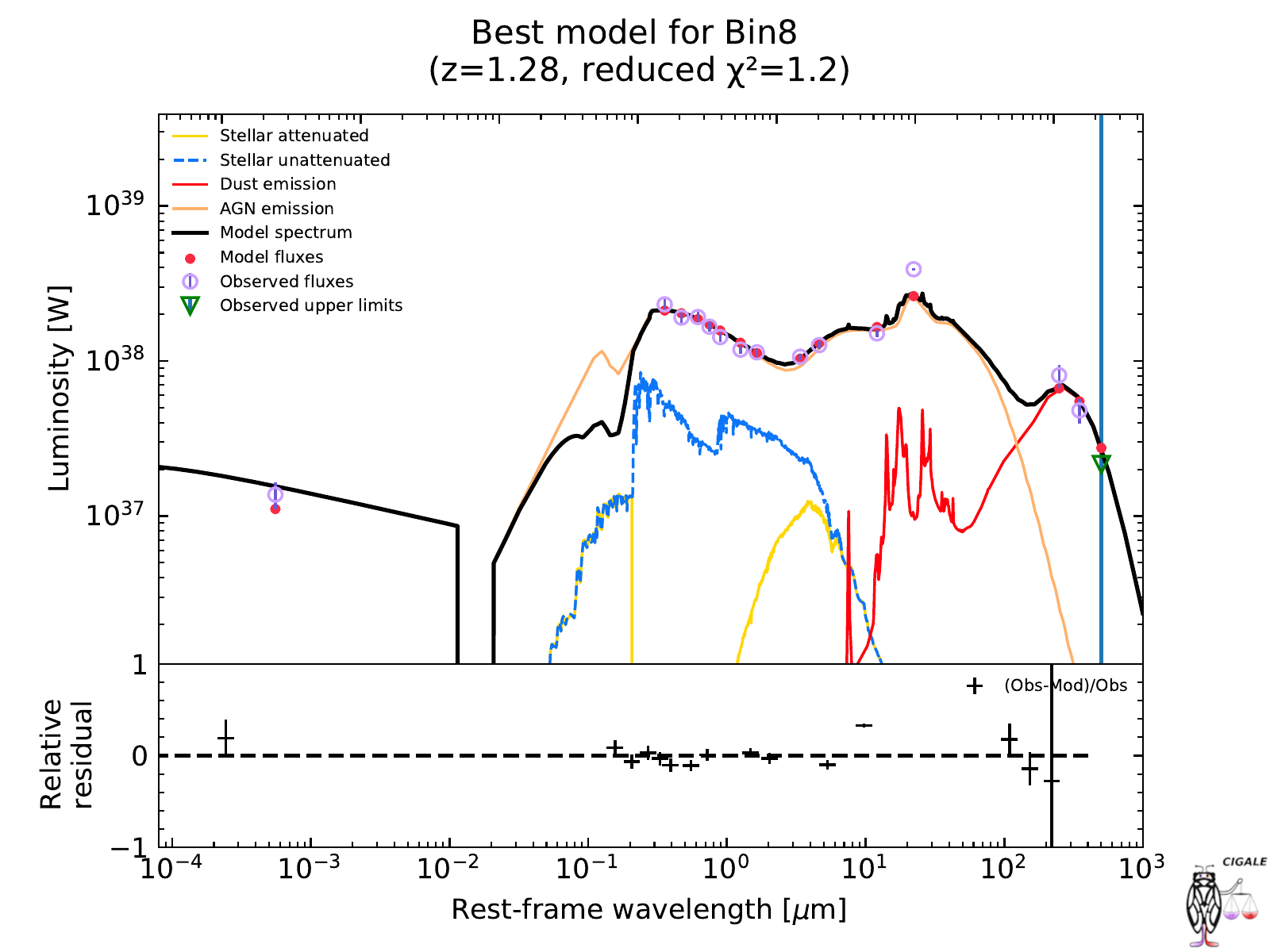}
    \includegraphics[width=0.33\textwidth]{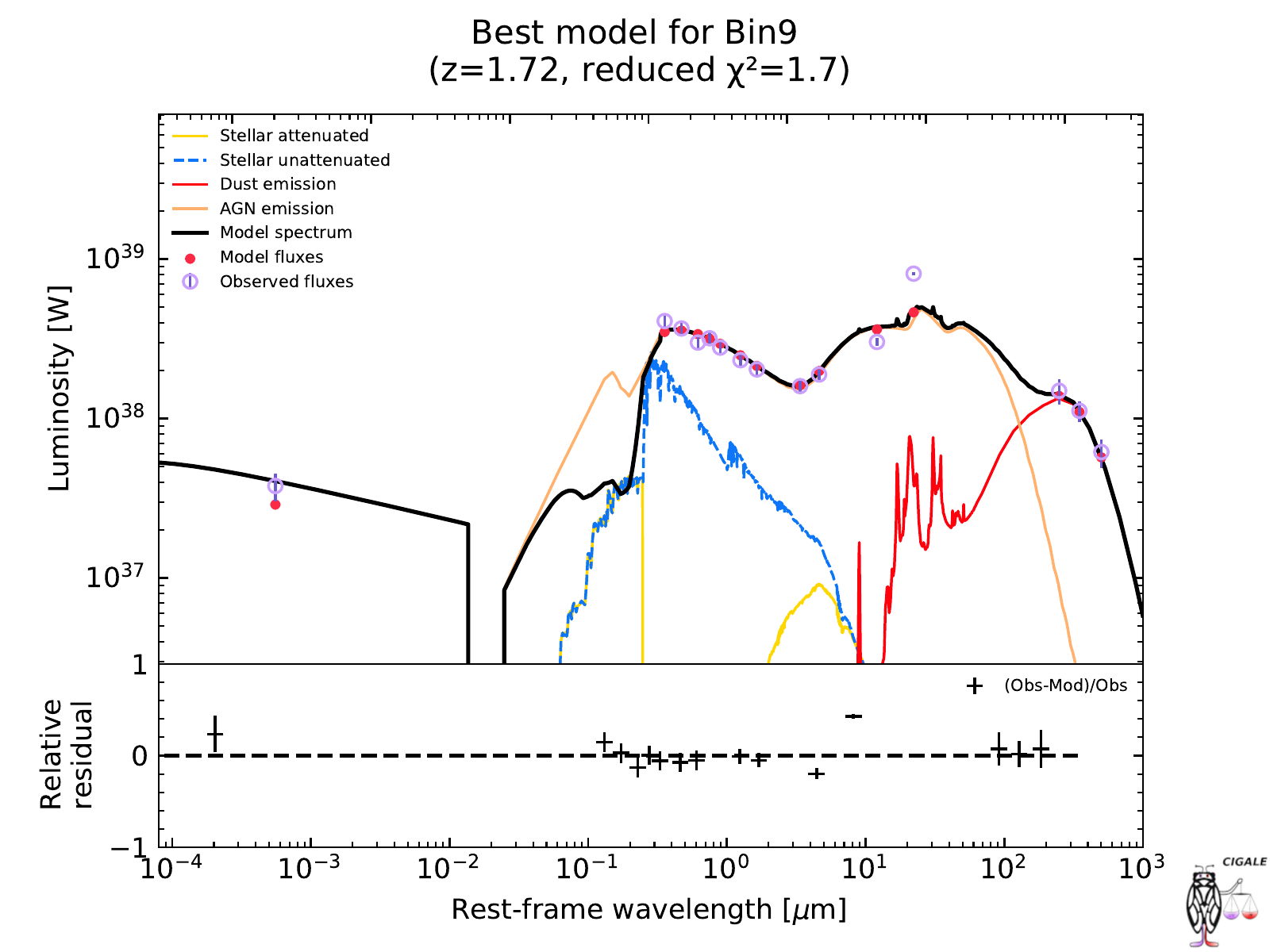}
    \includegraphics[width=0.33\textwidth]{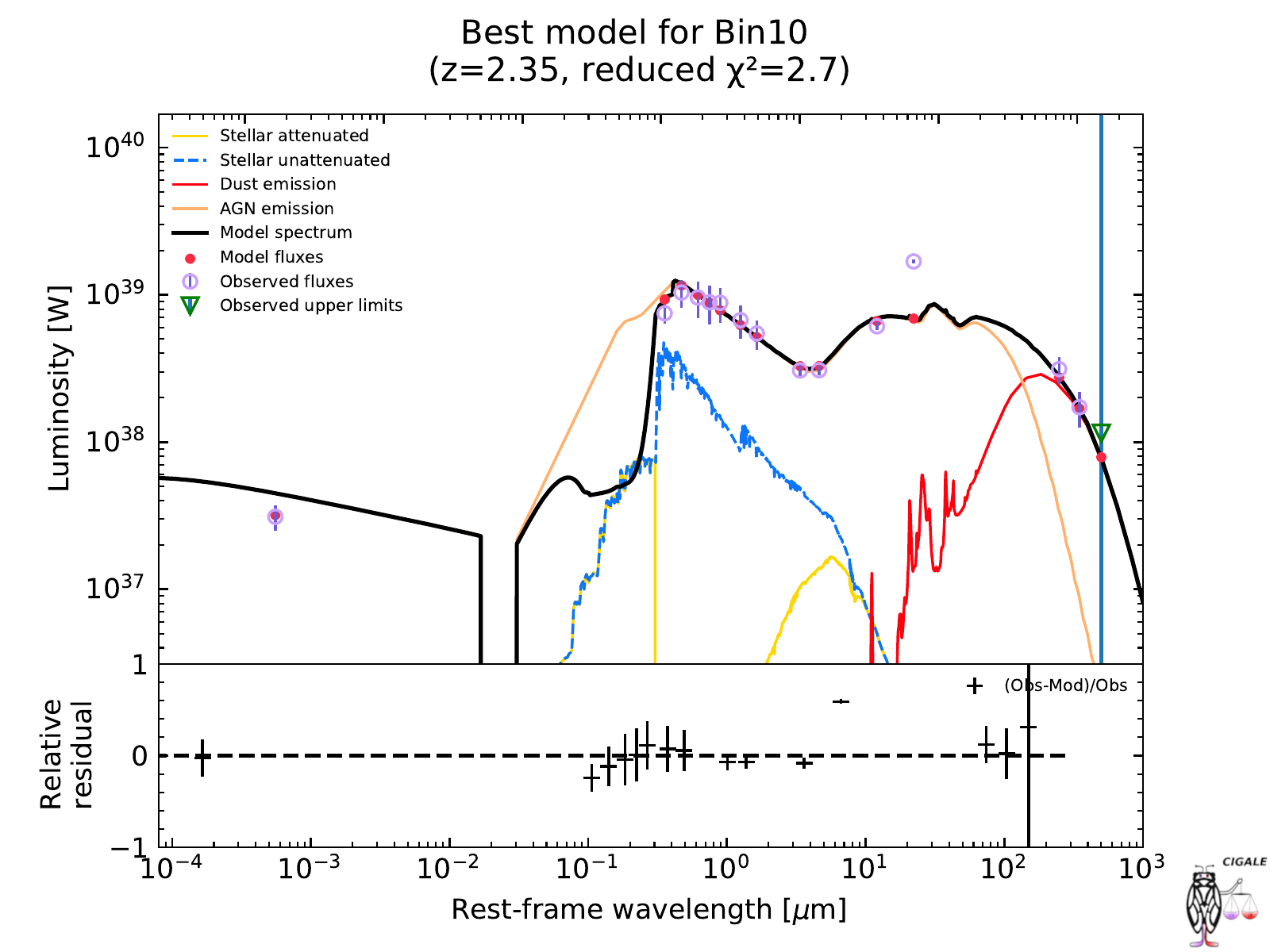}
    \caption{The SEDs created from using {\sc xCigale}. The photometry for each stack are the open purple circles, and the model predicted fluxes are the filled red circles. Different components are represented by different colors. \label{xcigfits}}
\end{figure*}
\begin{figure*}[ht]
\centering
\includegraphics[width=5.0in]{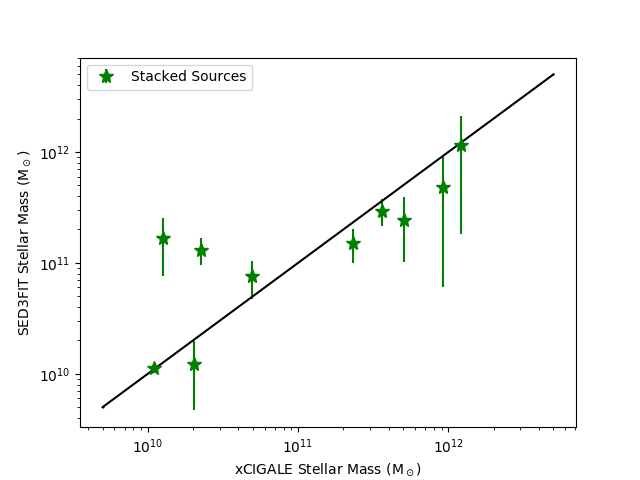}
\includegraphics[width=5.0in]{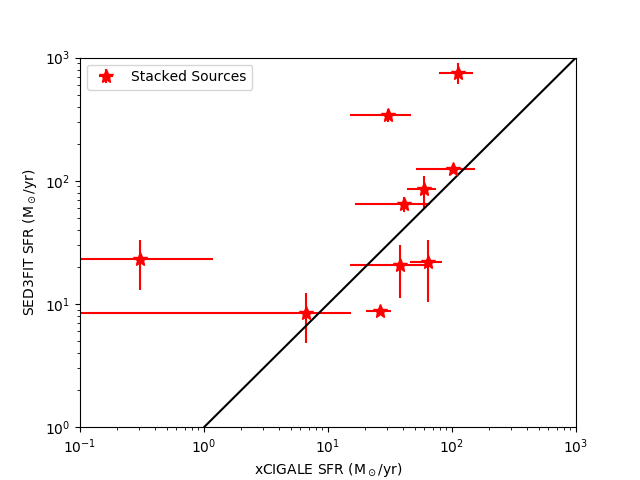}
\caption{A comparison between $M_\ast$ and SFR from two different SED fitting routines: {\sc Sed3Fit} and {\sc xCigale}. In general, $M_\ast$ is consistent among the routines. Overall, the SFR between the two routines is consistent as well. Using either set of results,  we see no dramatic change in the main sequence residuals (Figure \ref{resid}). \label{CIG_SED_Mstar}}
\end{figure*}

We compare the SFR, stellar mass, and dust mass parameters taken from {\sc xCigale} (Table 5) with the parameters taken from {\sc Sed3Fit}. We show all the {\sc xCigale} fits in Figure \ref{xcigfits}. For a visual comparison of the SFR and stellar mass, we plot the relationship between the two routine in Figure \ref{CIG_SED_Mstar}. On average, the stellar mass is consistent between the two routines. The same can be said for the majority of the SFRs. There are outliers present when comparing the two routines. To mitigate these, we also include the reduced chi squared value in both Table \ref{props} and Table 5. If we use the values from {\sc xCigale} instead and make a plot of Figure \ref{resid}, we see no dramatic change in the results (there is a simalar scatter, with stacks above, below, and on the main sequence). 

\begin{deluxetable}{lccccc}[h]
\tabletypesize{\footnotesize}
\tablecolumns{6}
\tablewidth{0pt}
\tablecaption{Stacked Parameters Using xCIGALE}
\tablehead{
\colhead{Bin} & \colhead{$M_\ast$} & \colhead{$M_{dust}$} & \colhead{SFR} & \colhead{Reduced $\chi^2$}\\
  & \colhead{($\log M_\odot$)} & \colhead{($\log M_\odot$)} & \colhead{($M_\odot /yr$)}}
\startdata
Bin 1 & 11.71  & 9.82  & 0.303 $\pm$ 0.874 & 3.339 \\ 
Bin 2 & 11.96  & 10.02  & 6.600 $\pm$ 8.586 &  0.655 \\
Bin 3 & 11.56  & 9.513  & 37.86 $\pm$ 22.82 &  0.574 \\
Bin 4 & 11.37  & 9.38  & 26.32 $\pm$ 5.90 & 1.263 \\
Bin 5 & 10.69  & 8.57  & 41.33 $\pm$ 24.86 & 0.956 \\
Bin 6 & 10.04  & 7.79  & 101.8 $\pm$ 50.11 & 1.046 \\
Bin 7 & 12.08  & 10.03 & 30.69 $\pm$ 15.74 & 0.265 \\
Bin 8 & 10.31  & 8.07  & 64.34 $\pm$ 18.169 & 1.521 \\
Bin 9 & 10.10 & 7.86  & 59.24 $\pm$ 15.57 & 1.692 \\
Bin 10 & 10.36 & 8.11  & 112.34 $\pm$ 34.06 & 2.674 \\
\enddata
\label{xcig}
\end{deluxetable}

\clearpage
\section{Data Availability}
The data underlying this article are available in \citet{2017ApJ...850...66A} [doi:10.3847/1538-4357/aa937ds] and \cite{2014ApJS..210...22V} [doi:10.1088/0067-0049/210/2/22]. The datasets were derived from sources in the public domain: XMM archive (\url{http://nxsa.esac.esa.int/nxsa-web/#home}), SDSS (\url{https://www.sdss.org/dr14/data_access/}), HerS (\url{http://hedam.lam.fr/HerMES/index/download}), and WISE (\url{https://irsa.ipac.caltech.edu/cgi-bin/Gator/nph-dd}).

\bibliographystyle{apj}
\bibliography{main_ref.bib}

\begin{thebibliography}{}
\expandafter\ifx\csname natexlab\endcsname\relax\def\natexlab#1{#1}\fi

\bibitem[{{Alexander} \& {Hickox}(2012)}]{2012NewAR..56...93A}
{Alexander}, D.~M., \& {Hickox}, R.~C. 2012, \nar, 56, 93

\bibitem[{{Ananna} {et~al.}(2017){Ananna}, {Salvato}, {LaMassa}, {Urry},
  {Cappelluti}, {Cardamone}, {Civano}, {Farrah}, {Gilfanov}, {Glikman},
  {Hamilton}, {Kirkpatrick}, {Lanzuisi}, {Marchesi}, {Merloni}, {Nandra},
  {Natarajan}, {Richards}, \& {Timlin}}]{2017ApJ...850...66A}
{Ananna}, T.~T., {Salvato}, M., {LaMassa}, S., {et~al.} 2017, \apj, 850, 66

\bibitem[{{Astropy Collaboration} {et~al.}(2013){Astropy Collaboration},
  {Robitaille}, {Tollerud}, {Greenfield}, {Droettboom}, {Bray}, {Aldcroft},
  {Davis}, {Ginsburg}, {Price-Whelan}, {Kerzendorf}, {Conley}, {Crighton},
  {Barbary}, {Muna}, {Ferguson}, {Grollier}, {Parikh}, {Nair}, {Unther},
  {Deil}, {Woillez}, {Conseil}, {Kramer}, {Turner}, {Singer}, {Fox}, {Weaver},
  {Zabalza}, {Edwards}, {Azalee Bostroem}, {Burke}, {Casey}, {Crawford},
  {Dencheva}, {Ely}, {Jenness}, {Labrie}, {Lim}, {Pierfederici}, {Pontzen},
  {Ptak}, {Refsdal}, {Servillat}, \& {Streicher}}]{2013A&A...558A..33A}
{Astropy Collaboration}, {Robitaille}, T.~P., {Tollerud}, E.~J., {et~al.} 2013,
  \aap, 558, A33

\bibitem[{{Astropy Collaboration} {et~al.}(2018){Astropy Collaboration},
  {Price-Whelan}, {Sip{\H{o}}cz}, {G{\"u}nther}, {Lim}, {Crawford}, {Conseil},
  {Shupe}, {Craig}, {Dencheva}, {Ginsburg}, {VanderPlas}, {Bradley},
  {P{\'e}rez-Su{\'a}rez}, {de Val-Borro}, {Aldcroft}, {Cruz}, {Robitaille},
  {Tollerud}, {Ardelean}, {Babej}, {Bach}, {Bachetti}, {Bakanov}, {Bamford},
  {Barentsen}, {Barmby}, {Baumbach}, {Berry}, {Biscani}, {Boquien}, {Bostroem},
  {Bouma}, {Brammer}, {Bray}, {Breytenbach}, {Buddelmeijer}, {Burke},
  {Calderone}, {Cano Rodr{\'\i}guez}, {Cara}, {Cardoso}, {Cheedella}, {Copin},
  {Corrales}, {Crichton}, {D'Avella}, {Deil}, {Depagne}, {Dietrich}, {Donath},
  {Droettboom}, {Earl}, {Erben}, {Fabbro}, {Ferreira}, {Finethy}, {Fox},
  {Garrison}, {Gibbons}, {Goldstein}, {Gommers}, {Greco}, {Greenfield},
  {Groener}, {Grollier}, {Hagen}, {Hirst}, {Homeier}, {Horton}, {Hosseinzadeh},
  {Hu}, {Hunkeler}, {Ivezi{\'c}}, {Jain}, {Jenness}, {Kanarek}, {Kendrew},
  {Kern}, {Kerzendorf}, {Khvalko}, {King}, {Kirkby}, {Kulkarni}, {Kumar},
  {Lee}, {Lenz}, {Littlefair}, {Ma}, {Macleod}, {Mastropietro}, {McCully},
  {Montagnac}, {Morris}, {Mueller}, {Mumford}, {Muna}, {Murphy}, {Nelson},
  {Nguyen}, {Ninan}, {N{\"o}the}, {Ogaz}, {Oh}, {Parejko}, {Parley}, {Pascual},
  {Patil}, {Patil}, {Plunkett}, {Prochaska}, {Rastogi}, {Reddy Janga},
  {Sabater}, {Sakurikar}, {Seifert}, {Sherbert}, {Sherwood-Taylor}, {Shih},
  {Sick}, {Silbiger}, {Singanamalla}, {Singer}, {Sladen}, {Sooley},
  {Sornarajah}, {Streicher}, {Teuben}, {Thomas}, {Tremblay}, {Turner},
  {Terr{\'o}n}, {van Kerkwijk}, {de la Vega}, {Watkins}, {Weaver}, {Whitmore},
  {Woillez}, {Zabalza}, \& {Astropy Contributors}}]{2018AJ....156..123A}
{Astropy Collaboration}, {Price-Whelan}, A.~M., {Sip{\H{o}}cz}, B.~M., {et~al.}
  2018, \aj, 156, 123

\bibitem[{{Azadi} {et~al.}(2017){Azadi}, {Coil}, {Aird}, {Reddy}, {Shapley},
  {Freeman}, {Kriek}, {Leung}, {Mobasher}, {Price}, {Sanders}, {Shivaei}, \&
  {Siana}}]{2017ApJ...835...27A}
{Azadi}, M., {Coil}, A.~L., {Aird}, J., {et~al.} 2017, \apj, 835, 27

\bibitem[{{Baldwin} {et~al.}(1981){Baldwin}, {Phillips}, \&
  {Terlevich}}]{1981PASP...93....5B}
{Baldwin}, J.~A., {Phillips}, M.~M., \& {Terlevich}, R. 1981, \pasp, 93, 5

\bibitem[{{Berta} {et~al.}(2013){Berta}, {Lutz}, {Santini}, {Wuyts}, {Rosario},
  {Brisbin}, {Cooray}, {Franceschini}, {Gruppioni}, {Hatziminaoglou}, {Hwang},
  {Le Floc'h}, {Magnelli}, {Nordon}, {Oliver}, {Page}, {Popesso}, {Pozzetti},
  {Pozzi}, {Riguccini}, {Rodighiero}, {Roseboom}, {Scott}, {Symeonidis},
  {Valtchanov}, {Viero}, \& {Wang}}]{2013A&A...551A.100B}
{Berta}, S., {Lutz}, D., {Santini}, P., {et~al.} 2013, \aap, 551, A100

\bibitem[{{Bertemes} {et~al.}(2018){Bertemes}, {Wuyts}, {Lutz}, {F{\"o}rster
  Schreiber}, {Genzel}, {Minchin}, {Mundell}, {Rosario}, {Saintonge}, \&
  {Tacconi}}]{2018MNRAS.478.1442B}
{Bertemes}, C., {Wuyts}, S., {Lutz}, D., {et~al.} 2018, \mnras, 478, 1442

\bibitem[{{Boquien} {et~al.}(2019){Boquien}, {Burgarella}, {Roehlly}, {Buat},
  {Ciesla}, {Corre}, {Inoue}, \& {Salas}}]{2019A&A...622A.103B}
{Boquien}, M., {Burgarella}, D., {Roehlly}, Y., {et~al.} 2019, \aap, 622, A103

\bibitem[{{Bruzual}(2007)}]{2007ASPC..374..303B}
{Bruzual}, G. 2007, in Astronomical Society of the Pacific Conference Series,
  Vol. 374, From Stars to Galaxies: Building the Pieces to Build Up the
  Universe, ed. A.~{Vallenari}, R.~{Tantalo}, L.~{Portinari}, \& A.~{Moretti},
  303

\bibitem[{{Bruzual} \& {Charlot}(2003)}]{bruzual2003}
{Bruzual}, G., \& {Charlot}, S. 2003, \mnras, 344, 1000

\bibitem[{{Burgarella} {et~al.}(2005){Burgarella}, {Buat}, \&
  {Iglesias-P{\'a}ramo}}]{2005MNRAS.360.1413B}
{Burgarella}, D., {Buat}, V., \& {Iglesias-P{\'a}ramo}, J. 2005, \mnras, 360,
  1413

\bibitem[{{Chabrier}(2003)}]{chabrier2003}
{Chabrier}, G. 2003, \pasp, 115, 763

\bibitem[{{Chang} {et~al.}(2018){Chang}, {Ferraro}, {Wang}, {Lim}, {Toba},
  {An}, {Chen}, {Smail}, {Shim}, {Ao}, {Bunker}, {Conselice}, {Cowley}, {da
  Cunha}, {Fan}, {Goto}, {Guo}, {Ho}, {Hwang}, {Lee}, {Lee}, {Micha{\l}owski},
  {Oteo}, {Scott}, {Serjeant}, {Shu}, {Simpson}, \&
  {Urquhart}}]{2018ApJ...865..103C}
{Chang}, Y.-Y., {Ferraro}, N., {Wang}, W.-H., {et~al.} 2018, \apj, 865, 103

\bibitem[{{Charlot} \& {Fall}(2000)}]{2000ApJ...539..718C}
{Charlot}, S., \& {Fall}, S.~M. 2000, \apj, 539, 718

\bibitem[{{Conroy} \& {White}(2013)}]{2013ApJ...762...70C}
{Conroy}, C., \& {White}, M. 2013, \apj, 762, 70

\bibitem[{{da Cunha} {et~al.}(2008){da Cunha}, {Charlot}, \&
  {Elbaz}}]{2008MNRAS.388.1595D}
{da Cunha}, E., {Charlot}, S., \& {Elbaz}, D. 2008, \mnras, 388, 1595

\bibitem[{{Elbaz} {et~al.}(2007){Elbaz}, {Daddi}, {Le Borgne}, {Dickinson},
  {Alexander}, {Chary}, {Starck}, {Brand t}, {Kitzbichler}, {MacDonald},
  {Nonino}, {Popesso}, {Stern}, \& {Vanzella}}]{2007A&A...468...33E}
{Elbaz}, D., {Daddi}, E., {Le Borgne}, D., {et~al.} 2007, \aap, 468, 33

\bibitem[{{Fliri} \& {Trujillo}(2016)}]{2016MNRAS.456.1359F}
{Fliri}, J., \& {Trujillo}, I. 2016, \mnras, 456, 1359

\bibitem[{{Fritz} {et~al.}(2006){Fritz}, {Franceschini}, \&
  {Hatziminaoglou}}]{fritz2006}
{Fritz}, J., {Franceschini}, A., \& {Hatziminaoglou}, E. 2006, \mnras, 366, 767

\bibitem[{{Glikman} {et~al.}(2004){Glikman}, {Gregg}, {Lacy}, {Helfand },
  {Becker}, \& {White}}]{2004ApJ...607...60G}
{Glikman}, E., {Gregg}, M.~D., {Lacy}, M., {et~al.} 2004, \apj, 607, 60

\bibitem[{{Glikman} {et~al.}(2012){Glikman}, {Urrutia}, {Lacy}, {Djorgovski},
  {Mahabal}, {Myers}, {Ross}, {Petitjean}, {Ge}, {Schneider}, \&
  {York}}]{2012ApJ...757...51G}
{Glikman}, E., {Urrutia}, T., {Lacy}, M., {et~al.} 2012, \apj, 757, 51

\bibitem[{{Groves} {et~al.}(2015){Groves}, {Schinnerer}, {Leroy}, {Galametz},
  {Walter}, {Bolatto}, {Hunt}, {Dale}, {Calzetti}, {Croxall}, \&
  {Kennicutt}}]{groves2015}
{Groves}, B.~A., {Schinnerer}, E., {Leroy}, A., {et~al.} 2015, \apj, 799, 96

\bibitem[{{Hainline} {et~al.}(2011){Hainline}, {Shapley}, {Greene}, \&
  {Steidel}}]{2011ApJ...733...31H}
{Hainline}, K.~N., {Shapley}, A.~E., {Greene}, J.~E., \& {Steidel}, C.~C. 2011,
  \apj, 733, 31

\bibitem[{{Hickox} {et~al.}(2009){Hickox}, {Jones}, {Forman}, {Murray},
  {Kochanek}, {Eisenstein}, {Jannuzi}, {Dey}, {Brown}, {Stern}, {Eisenhardt},
  {Gorjian}, {Brodwin}, {Narayan}, {Cool}, {Kenter}, {Caldwell}, \&
  {Anderson}}]{2009ApJ...696..891H}
{Hickox}, R.~C., {Jones}, C., {Forman}, W.~R., {et~al.} 2009, \apj, 696, 891

\bibitem[{{Hopkins} {et~al.}(2007){Hopkins}, {Richards}, \&
  {Hernquist}}]{2007ApJ...654..731H}
{Hopkins}, P.~F., {Richards}, G.~T., \& {Hernquist}, L. 2007, \apj, 654, 731

\bibitem[{{Hopkins} {et~al.}(2006){Hopkins}, {Somerville}, {Hernquist}, {Cox},
  {Robertson}, \& {Li}}]{2006ApJ...652..864H}
{Hopkins}, P.~F., {Somerville}, R.~S., {Hernquist}, L., {et~al.} 2006, \apj,
  652, 864

\bibitem[{{Hunter}(2007)}]{2007CSE.....9...90H}
{Hunter}, J.~D. 2007, Computing in Science and Engineering, 9, 90

\bibitem[{{Imanishi} {et~al.}(2016){Imanishi}, {Nakanishi}, \&
  {Izumi}}]{2016ApJ...822L..10I}
{Imanishi}, M., {Nakanishi}, K., \& {Izumi}, T. 2016, \apjl, 822, L10

\bibitem[{{Jiang} {et~al.}(2014){Jiang}, {Fan}, {Bian}, {McGreer}, {Strauss},
  {Annis}, {Buck}, {Green}, {Hodge}, {Myers}, {Rafiee}, \&
  {Richards}}]{2014ApJS..213...12J}
{Jiang}, L., {Fan}, X., {Bian}, F., {et~al.} 2014, \apjs, 213, 12

\bibitem[{{Just} {et~al.}(2007){Just}, {Brandt}, {Shemmer}, {Steffen},
  {Schneider}, {Chartas}, \& {Garmire}}]{2007ApJ...665.1004J}
{Just}, D.~W., {Brandt}, W.~N., {Shemmer}, O., {et~al.} 2007, \apj, 665, 1004

\bibitem[{{Kartaltepe} {et~al.}(2012){Kartaltepe}, {Dickinson}, {Alexander},
  {Bell}, {Dahlen}, {Elbaz}, {Faber}, {Lotz}, {McIntosh}, {Wiklind}, {Altieri},
  {Aussel}, {Bethermin}, {Bournaud}, {Charmand aris}, {Conselice}, {Cooray},
  {Dannerbauer}, {Dav{\'e}}, {Dunlop}, {Dekel}, {Ferguson}, {Grogin}, {Hwang},
  {Ivison}, {Kocevski}, {Koekemoer}, {Koo}, {Lai}, {Leiton}, {Lucas}, {Lutz},
  {Magdis}, {Magnelli}, {Morrison}, {Mozena}, {Mullaney}, {Newman}, {Pope},
  {Popesso}, {van der Wel}, {Weiner}, \& {Wuyts}}]{2012ApJ...757...23K}
{Kartaltepe}, J.~S., {Dickinson}, M., {Alexander}, D.~M., {et~al.} 2012, \apj,
  757, 23

\bibitem[{{Kirkpatrick} {et~al.}(2019){Kirkpatrick}, {Sharon}, {Keller}, \&
  {Pope}}]{kirkpatrick2019}
{Kirkpatrick}, A., {Sharon}, C., {Keller}, E., \& {Pope}, A. 2019, \apj, 879,
  41

\bibitem[{{Kirkpatrick} {et~al.}(2012){Kirkpatrick}, {Pope}, {Alexander},
  {Charmandaris}, {Daddi}, {Dickinson}, {Elbaz}, {Gabor}, {Hwang}, {Ivison},
  {Mullaney}, {Pannella}, {Scott}, {Altieri}, {Aussel}, {Bournaud}, {Buat},
  {Coia}, {Dannerbauer}, {Dasyra}, {Kartaltepe}, {Leiton}, {Lin}, {Magdis},
  {Magnelli}, {Morrison}, {Popesso}, \& {Valtchanov}}]{kirkpatrick2012}
{Kirkpatrick}, A., {Pope}, A., {Alexander}, D.~M., {et~al.} 2012, \apj, 759,
  139

\bibitem[{{Kirkpatrick} {et~al.}(2017){Kirkpatrick}, {Pope}, {Sajina}, {Dale},
  {D{\'\i}az-Santos}, {Hayward}, {Shi}, {Somerville}, {Stierwalt}, {Armus},
  {Kartaltepe}, {Kocevski}, {McIntosh}, {Sanders}, \&
  {Yan}}]{2017ApJ...843...71K}
{Kirkpatrick}, A., {Pope}, A., {Sajina}, A., {et~al.} 2017, \apj, 843, 71

\bibitem[{{Kirkpatrick} {et~al.}(2020){Kirkpatrick}, {Urry}, {Brewster},
  {Cooke}, {Estrada}, {Glikman}, {Hamblin}, {Ananna}, {Carlile}, {Coleman},
  {Johnson}, {Kartaltepe}, {LaMassa}, {Marchesi}, {Powell}, {Sanders},
  {Treister}, \& {Jan Turner}}]{kirkpatrick2020}
{Kirkpatrick}, A., {Urry}, C.~M., {Brewster}, J., {et~al.} 2020, \apj, 900, 5

\bibitem[{{La Plante} \& {Trac}(2016)}]{2016ApJ...828...90L}
{La Plante}, P., \& {Trac}, H. 2016, \apj, 828, 90

\bibitem[{{LaMassa} {et~al.}(2013{\natexlab{a}}){LaMassa}, {Urry}, {Glikman},
  {Cappelluti}, {Civano}, {Comastri}, {Treister}, {B{\"o}hringer}, {Cardamone},
  {Chon}, {Kephart}, {Murray}, {Richards}, {Ross}, {Rozner}, \&
  {Schawinski}}]{2013MNRAS.432.1351L}
{LaMassa}, S.~M., {Urry}, C.~M., {Glikman}, E., {et~al.} 2013{\natexlab{a}},
  \mnras, 432, 1351

\bibitem[{{LaMassa} {et~al.}(2013{\natexlab{b}}){LaMassa}, {Urry},
  {Cappelluti}, {Civano}, {Ranalli}, {Glikman}, {Treister}, {Richards},
  {Ballantyne}, {Stern}, {Comastri}, {Cardamone}, {Schawinski},
  {B{\"o}hringer}, {Chon}, {Murray}, {Green}, \&
  {Nandra}}]{2013MNRAS.436.3581L}
{LaMassa}, S.~M., {Urry}, C.~M., {Cappelluti}, N., {et~al.} 2013{\natexlab{b}},
  \mnras, 436, 3581

\bibitem[{{LaMassa} {et~al.}(2015){LaMassa}, {Cales}, {Moran}, {Myers},
  {Richards}, {Eracleous}, {Heckman}, {Gallo}, \& {Urry}}]{2015ApJ...800..144L}
{LaMassa}, S.~M., {Cales}, S., {Moran}, E.~C., {et~al.} 2015, \apj, 800, 144

\bibitem[{{LaMassa} {et~al.}(2016){LaMassa}, {Urry}, {Cappelluti},
  {B{\"o}hringer}, {Comastri}, {Glikman}, {Richards}, {Ananna}, {Brusa},
  {Cardamone}, {Chon}, {Civano}, {Farrah}, {Gilfanov}, {Green}, {Komossa},
  {Lira}, {Makler}, {Marchesi}, {Pecoraro}, {Ranalli}, {Salvato}, {Schawinski},
  {Stern}, {Treister}, \& {Viero}}]{2016ApJ...817..172L}
{LaMassa}, S.~M., {Urry}, C.~M., {Cappelluti}, N., {et~al.} 2016, \apj, 817,
  172

\bibitem[{{Lusso} \& {Risaliti}(2017)}]{2017A&A...602A..79L}
{Lusso}, E., \& {Risaliti}, G. 2017, \aap, 602, A79

\bibitem[{{Lutz} {et~al.}(2004){Lutz}, {Maiolino}, {Spoon}, \&
  {Moorwood}}]{2004A&A...418..465L}
{Lutz}, D., {Maiolino}, R., {Spoon}, H.~W.~W., \& {Moorwood}, A.~F.~M. 2004,
  \aap, 418, 465

\bibitem[{{Lyu} \& {Rieke}(2017)}]{lyu2017}
{Lyu}, J., \& {Rieke}, G.~H. 2017, \apj, 841, 76

\bibitem[{{McKinney} {et~al.}(2021){McKinney}, {Hayward}, {Rosenthal},
  {Martinez-Galarza}, {Pope}, {Sajina}, \& {Smith}}]{mckinney2021}
{McKinney}, J., {Hayward}, C.~C., {Rosenthal}, L.~J., {et~al.} 2021, arXiv
  e-prints, arXiv:2103.12747

\bibitem[{{McMahon} {et~al.}(2013){McMahon}, {Banerji}, {Gonzalez}, {Koposov},
  {Bejar}, {Lodieu}, {Rebolo}, \& {VHS Collaboration}}]{2013Msngr.154...35M}
{McMahon}, R.~G., {Banerji}, M., {Gonzalez}, E., {et~al.} 2013, The Messenger,
  154, 35

\bibitem[{{Mullaney} {et~al.}(2011){Mullaney}, {Alexander}, {Goulding}, \&
  {Hickox}}]{2011MNRAS.414.1082M}
{Mullaney}, J.~R., {Alexander}, D.~M., {Goulding}, A.~D., \& {Hickox}, R.~C.
  2011, \mnras, 414, 1082

\bibitem[{{Mullaney} {et~al.}(2015){Mullaney}, {Alexander}, {Aird}, {Bernhard},
  {Daddi}, {Del Moro}, {Dickinson}, {Elbaz}, {Harrison}, {Juneau}, {Liu},
  {Pannella}, {Rosario}, {Santini}, {Sargent}, {Schreiber}, {Simpson}, \&
  {Stanley}}]{2015MNRAS.453L..83M}
{Mullaney}, J.~R., {Alexander}, D.~M., {Aird}, J., {et~al.} 2015, \mnras, 453,
  L83

\bibitem[{{Noeske} {et~al.}(2007){Noeske}, {Weiner}, {Faber}, {Papovich},
  {Koo}, {Somerville}, {Bundy}, {Conselice}, {Newman}, {Schiminovich}, {Le
  Floc'h}, {Coil}, {Rieke}, {Lotz}, {Primack}, {Barmby}, {Cooper}, {Davis},
  {Ellis}, {Fazio}, {Guhathakurta}, {Huang}, {Kassin}, {Martin}, {Phillips},
  {Rich}, {Small}, {Willmer}, \& {Wilson}}]{2007ApJ...660L..43N}
{Noeske}, K.~G., {Weiner}, B.~J., {Faber}, S.~M., {et~al.} 2007, \apjl, 660,
  L43

\bibitem[{{Noll} {et~al.}(2009){Noll}, {Burgarella}, {Giovannoli}, {Buat},
  {Marcillac}, \& {Mu{\~n}oz-Mateos}}]{2009A&A...507.1793N}
{Noll}, S., {Burgarella}, D., {Giovannoli}, E., {et~al.} 2009, \aap, 507, 1793

\bibitem[{{Papovich} {et~al.}(2016){Papovich}, {Shipley}, {Mehrtens}, {Lanham},
  {Lacy}, {Ciardullo}, {Finkelstein}, {Bassett}, {Behroozi}, {Blanc}, {de
  Jong}, {DePoy}, {Drory}, {Gawiser}, {Gebhardt}, {Gronwall}, {Hill}, {Hopp},
  {Jogee}, {Kawinwanichakij}, {Marshall}, {McLinden}, {Mentuch Cooper},
  {Somerville}, {Steinmetz}, {Tran}, {Tuttle}, {Viero}, {Wechsler}, \&
  {Zeimann}}]{2016ApJS..224...28P}
{Papovich}, C., {Shipley}, H.~V., {Mehrtens}, N., {et~al.} 2016, \apjs, 224, 28

\bibitem[{{Perez} \& {Granger}(2007)}]{2007CSE.....9c..21P}
{Perez}, F., \& {Granger}, B.~E. 2007, Computing in Science and Engineering, 9,
  21

\bibitem[{{Perna} {et~al.}(2018){Perna}, {Sargent}, {Brusa}, {Daddi},
  {Feruglio}, {Cresci}, {Lanzuisi}, {Lusso}, {Comastri}, {Coogan}, {D'Amato},
  {Gilli}, {Piconcelli}, \& {Vignali}}]{perna2018}
{Perna}, M., {Sargent}, M.~T., {Brusa}, M., {et~al.} 2018, \aap, 619, A90

\bibitem[{{Ricci} {et~al.}(2017){Ricci}, {Bauer}, {Treister}, {Schawinski},
  {Privon}, {Blecha}, {Arevalo}, {Armus}, {Harrison}, {Ho}, {Iwasawa},
  {Sanders}, \& {Stern}}]{2017MNRAS.468.1273R}
{Ricci}, C., {Bauer}, F.~E., {Treister}, E., {et~al.} 2017, \mnras, 468, 1273

\bibitem[{{Richards} {et~al.}(2006){Richards}, {Lacy}, {Storrie-Lombardi},
  {Hall}, {Gallagher}, {Hines}, {Fan}, {Papovich}, {Vanden Berk}, {Trammell},
  {Schneider}, {Vestergaard}, {York}, {Jester}, {Anderson}, {Budav{\'a}ri}, \&
  {Szalay}}]{2006ApJS..166..470R}
{Richards}, G.~T., {Lacy}, M., {Storrie-Lombardi}, L.~J., {et~al.} 2006, \apjs,
  166, 470

\bibitem[{{Sanders} {et~al.}(1988){Sanders}, {Soifer}, {Elias}, {Neugebauer},
  \& {Matthews}}]{1988ApJ...328L..35S}
{Sanders}, D.~B., {Soifer}, B.~T., {Elias}, J.~H., {Neugebauer}, G., \&
  {Matthews}, K. 1988, \apjl, 328, L35

\bibitem[{{Schreiber} {et~al.}(2015){Schreiber}, {Pannella}, {Elbaz},
  {B{\'e}thermin}, {Inami}, {Dickinson}, {Magnelli}, {Wang}, {Aussel}, {Daddi},
  {Juneau}, {Shu}, {Sargent}, {Buat}, {Faber}, {Ferguson}, {Giavalisco},
  {Koekemoer}, {Magdis}, {Morrison}, {Papovich}, {Santini}, \&
  {Scott}}]{2015A&A...575A..74S}
{Schreiber}, C., {Pannella}, M., {Elbaz}, D., {et~al.} 2015, \aap, 575, A74

\bibitem[{{Schulze} {et~al.}(2019){Schulze}, {Silverman}, {Daddi},
  {Rujopakarn}, {Liu}, {Schramm}, {Mainieri}, {Imanishi}, {Hirschmann}, \&
  {Jahnke}}]{2019MNRAS.488.1180S}
{Schulze}, A., {Silverman}, J.~D., {Daddi}, E., {et~al.} 2019, \mnras, 488,
  1180

\bibitem[{{Scoville} {et~al.}(2016){Scoville}, {Sheth}, {Aussel}, {Vanden
  Bout}, {Capak}, {Bongiorno}, {Casey}, {Murchikova}, {Koda},
  {{\'A}lvarez-M{\'a}rquez}, {Lee}, {Laigle}, {McCracken}, {Ilbert}, {Pope},
  {Sanders}, {Chu}, {Toft}, {Ivison}, \& {Manohar}}]{scoville2016}
{Scoville}, N., {Sheth}, K., {Aussel}, H., {et~al.} 2016, \apj, 820, 83

\bibitem[{{Serra} {et~al.}(2011){Serra}, {Amblard}, {Temi}, {Burgarella},
  {Giovannoli}, {Buat}, {Noll}, \& {Im}}]{2011ApJ...740...22S}
{Serra}, P., {Amblard}, A., {Temi}, P., {et~al.} 2011, \apj, 740, 22

\bibitem[{{Stacey} {et~al.}(2020){Stacey}, {McKean}, {Powell}, {Vegetti},
  {Rizzo}, {Spingola}, {Auger}, {Ivison}, \& {van der
  Werf}}]{2020arXiv200901277S}
{Stacey}, H.~R., {McKean}, J.~P., {Powell}, D.~M., {et~al.} 2020, arXiv
  e-prints, arXiv:2009.01277

\bibitem[{{Stanley} {et~al.}(2017){Stanley}, {Alexander}, {Harrison},
  {Rosario}, {Wang}, {Aird}, {Bourne}, {Dunne}, {Dye}, {Eales}, {Knudsen},
  {Micha{\l}owski}, {Valiante}, {De Zotti}, {Furlanetto}, {Ivison}, {Maddox},
  \& {Smith}}]{2017MNRAS.472.2221S}
{Stanley}, F., {Alexander}, D.~M., {Harrison}, C.~M., {et~al.} 2017, \mnras,
  472, 2221

\bibitem[{{Steffen} {et~al.}(2006){Steffen}, {Strateva}, {Brandt}, {Alexander},
  {Koekemoer}, {Lehmer}, {Schneider}, \& {Vignali}}]{2006AJ....131.2826S}
{Steffen}, A.~T., {Strateva}, I., {Brandt}, W.~N., {et~al.} 2006, \aj, 131,
  2826

\bibitem[{{Stern}(2015)}]{2015ApJ...807..129S}
{Stern}, D. 2015, \apj, 807, 129

\bibitem[{{Stern} {et~al.}(2012){Stern}, {Assef}, {Benford}, {Blain}, {Cutri},
  {Dey}, {Eisenhardt}, {Griffith}, {Jarrett}, {Lake}, {Masci}, {Petty},
  {Stanford}, {Tsai}, {Wright}, {Yan}, {Harrison}, \&
  {Madsen}}]{2012ApJ...753...30S}
{Stern}, D., {Assef}, R.~J., {Benford}, D.~J., {et~al.} 2012, \apj, 753, 30

\bibitem[{{Symeonidis} {et~al.}(2016){Symeonidis}, {Giblin}, {Page}, {Pearson},
  {Bendo}, {Seymour}, \& {Oliver}}]{symeonidis2016}
{Symeonidis}, M., {Giblin}, B.~M., {Page}, M.~J., {et~al.} 2016, \mnras, 459,
  257

\bibitem[{{Tadhunter} {et~al.}(2007){Tadhunter}, {Dicken}, {Holt}, {Inskip},
  {Morganti}, {Axon}, {Buchanan}, {Gonz{\'a}lez Delgado}, {Barthel}, \& {van
  Bemmel}}]{tadhunter2007}
{Tadhunter}, C., {Dicken}, D., {Holt}, J., {et~al.} 2007, \apjl, 661, L13

\bibitem[{{Timlin} {et~al.}(2016){Timlin}, {Ross}, {Richards}, {Lacy}, {Ryan},
  {Stone}, {Bauer}, {Brandt}, {Fan}, {Glikman}, {Haggard}, {Jiang}, {LaMassa},
  {Lin}, {Makler}, {McGehee}, {Myers}, {Schneider}, {Urry}, {Wollack}, \&
  {Zakamska}}]{2016ApJS..225....1T}
{Timlin}, J.~D., {Ross}, N.~P., {Richards}, G.~T., {et~al.} 2016, \apjs, 225, 1

\bibitem[{{Treister} {et~al.}(2010){Treister}, {Natarajan}, {Sanders}, {Urry},
  {Schawinski}, \& {Kartaltepe}}]{2010Sci...328..600T}
{Treister}, E., {Natarajan}, P., {Sanders}, D.~B., {et~al.} 2010, Science, 328,
  600

\bibitem[{{Twite} {et~al.}(2012){Twite}, {Conselice}, {Buitrago}, {Noeske},
  {Weiner}, {Acosta-Pulido}, \& {Bauer}}]{2012MNRAS.420.1061T}
{Twite}, J.~W., {Conselice}, C.~J., {Buitrago}, F., {et~al.} 2012, \mnras, 420,
  1061

\bibitem[{{Urrutia} {et~al.}(2012){Urrutia}, {Lacy}, {Spoon}, {Glikman},
  {Petric}, \& {Schulz}}]{2012ApJ...757..125U}
{Urrutia}, T., {Lacy}, M., {Spoon}, H., {et~al.} 2012, \apj, 757, 125

\bibitem[{{van der Walt} {et~al.}(2011){van der Walt}, {Colbert}, \&
  {Varoquaux}}]{2011CSE....13b..22V}
{van der Walt}, S., {Colbert}, S.~C., \& {Varoquaux}, G. 2011, Computing in
  Science and Engineering, 13, 22

\bibitem[{{Viero} {et~al.}(2014){Viero}, {Asboth}, {Roseboom}, {Moncelsi},
  {Marsden}, {Mentuch Cooper}, {Zemcov}, {Addison}, {Baker}, {Beelen}, {Bock},
  {Bridge}, {Conley}, {Devlin}, {Dor{\'e}}, {Farrah}, {Finkelstein},
  {Font-Ribera}, {Geach}, {Gebhardt}, {Gill}, {Glenn}, {Hajian}, {Halpern},
  {Jogee}, {Kurczynski}, {Lapi}, {Negrello}, {Oliver}, {Papovich}, {Quadri},
  {Ross}, {Scott}, {Schulz}, {Somerville}, {Spergel}, {Vieira}, {Wang}, \&
  {Wechsler}}]{2014ApJS..210...22V}
{Viero}, M.~P., {Asboth}, V., {Roseboom}, I.~G., {et~al.} 2014, \apjs, 210, 22

\bibitem[{{Virtanen} {et~al.}(2020){Virtanen}, {Gommers}, {Burovski},
  {Oliphant}, {Weckesser}, {Cournapeau}, {Alexbrc}, {Peterson}, {Reddy},
  {Wilson}, {Haberland}, {Mayorov}, {Endolith}, {Nelson}, {Der Van Walt},
  {Laxalde}, {Brett}, {Polat}, {Larson}, {Millman}, {Lars}, {Van Mulbregt},
  {Eric-Jones}, {Carey}, {Moore}, {Kern}, {Leslie}, {Perktold}, {Striega}, \&
  {Feng}}]{2020zndo....595738V}
{Virtanen}, P., {Gommers}, R., {Burovski}, E., {et~al.} 2020, {scipy/scipy:
  SciPy 1.5.3}, doi:10.5281/zenodo.595738

\bibitem[{{Wang} {et~al.}(2015){Wang}, {Viero}, {Ross}, {Asboth},
  {B{\'e}thermin}, {Bock}, {Clements}, {Conley}, {Cooray}, {Farrah}, {Hajian},
  {Han}, {Lagache}, {Marsden}, {Myers}, {Norberg}, {Oliver}, {Page},
  {Symeonidis}, {Schulz}, {Wang}, \& {Zemcov}}]{wang2015}
{Wang}, L., {Viero}, M., {Ross}, N.~P., {et~al.} 2015, \mnras, 449, 4476

\bibitem[{{Xie} {et~al.}(2021){Xie}, {Ho}, {Zhuang}, \&
  {Shangguan}}]{2021ApJ...910..124X}
{Xie}, Y., {Ho}, L.~C., {Zhuang}, M.-Y., \& {Shangguan}, J. 2021, \apj, 910,
  124

\bibitem[{{Yang} {et~al.}(2020){Yang}, {Boquien}, {Buat}, {Burgarella},
  {Ciesla}, {Duras}, {Stalevski}, {Brandt}, \&
  {Papovich}}]{2020MNRAS.491..740Y}
{Yang}, G., {Boquien}, M., {Buat}, V., {et~al.} 2020, \mnras, 491, 740

\end{thebibliography}

\end{document}